\documentclass[journal]{IEEEtran}
\IEEEoverridecommandlockouts
\usepackage{stfloats}
\usepackage{cite}
\usepackage{amsmath,amssymb,amsfonts}
\usepackage{algorithmic}
\usepackage{algorithm}
\usepackage{graphicx}
\usepackage{textcomp}
\usepackage{xcolor}
\usepackage{url}
\usepackage[short]{optidef}
\usepackage{cuted}
\usepackage{multirow}
\makeatletter
\newcommand{\biggg}{\bBigg@\thr@@}
\newcommand{\Biggg}{\bBigg@{3.5}}

\makeatother

\IEEEoverridecommandlockouts
\newtheorem{thm}{Theorem}[section]

\newtheorem{prop}[thm]{Proposition}

\newtheorem{defn}{Definition}

\newtheorem{rem}{Remark}

\def\BibTeX{{\rm B\kern-.05em{\sc i\kern-.025em b}\kern-.08em
    T\kern-.1667em\lower.7ex\hbox{E}\kern-.125emX}}
\begin{document} \sloppy

\title{On the Application of Uplink/Downlink Decoupled Access in Heterogeneous Mobile Edge Computing}

\author{Yao Shi, \IEEEmembership{Member, IEEE},\ Emad Alsusa, \IEEEmembership{Senior Member, IEEE}\ and Mohammed W. Baidas, \IEEEmembership{Senior Member, IEEE}
\thanks{Yao Shi is with the School of Electronic and Information Engineering, Harbin Institute of Technology, Shenzhen, China (e-mail: shiyao@hit.edu.cn).}
\thanks{Emad Alsusa is with the School of Electrical and Electronic Engineering, University of Manchester, Manchester M1 3WE, UK (e-mail: yao.shi@manchester.ac.uk, e.alsusa@manchester.ac.uk). }
\thanks{Mohammed W. Baidas is with the Electrical Engineering Department, College of Engineering and Petroleum, Kuwait University, Kuwait (e-mail: m.baidas@ku.edu.kw).}}

\maketitle

\begin{abstract}
Mobile edge computing (MEC) is a key player in low latency 5G networks with the task to resolve the conflict between computationally-intensive mobile applications and resource-limited mobile devices (MDs). As such, there has been intense interest in this topic, especially in multi-user single-server and homogeneous multi-server scenarios. However, the research in the heterogeneous multi-server scenario is limited, where the servers are located at small base-stations (SBSs), macro base-stations (MBSs), or the cloud with different computing and communication capabilities. On the other hand, computational-tasks offloading is limited by the type of MD-BS association with almost all previous works focusing on offloading the MD's computational tasks to the MEC servers/cloudlets at its serving BS. However, in multi-BS association, or downlink/uplink decoupled (DUDe) scenarios, an MD can be served by multiple BSs and hence has multiple offloading choices. Motivated by this, we proposed a joint BS association and subchannel allocation algorithm based on a student-project allocation (SPA) matching approach to minimize the network sum-latency, which break the constraint that one MD must connect to the same BS in the UL and DL, and jointly consider the communication and computational disparity of SBS and MBS cloudlets in heterogeneous MEC networks. Moreover, an optimal power allocation scheme is proposed to optimize the system performance subject to the predefined quality of service constraints. Our results show that the proposed scheme is superior to benchmark techniques in enabling effective use of the computational and communication resources in heterogeneous MEC networks.

\end{abstract}


\section{Introduction}
As mobile devices (MDs) became an indispensable part of daily life, numerous applications, from language recognition and map navigation to augmented reality and cloud gaming, to name a few, came to life. Such applications, as well as forthcoming ones, tend to be computation-intensive and latency-sensitive. Due to battery power and computation limitations in MDs, computationally-intensive tasks and workloads will often need to be offloaded to powerful servers, under the notion of cloud computing. To further reduce latency and avoid long-distance transmissions delays from the MDs to the cloud servers, mobile edge computing (MEC) has been proposed as a booster technology to cloud computing, where edge servers are deployed at the cellular base-stations (BS), such as macro base-stations (MBSs) and/or small base-stations (SBSs). With the development of  ultra-dense 5G networks, MEC has become a highly desirable technique to minimize latency while satisfying the capacity requirements of mobile communications and Internet of Things (IoT) devices which may aggregate data from multiple industrial sensors to transfer to the MEC for digital-twinning purposes, and/or processing and feedback control signalling for actuation \cite{9123504,mec23}.

Unlike cloud computing, MEC is limited by real-time delay constraints and computing resources \cite{mec2}, hence if too many tasks are offloaded to one MEC, severe congestion and latency may occur. Also, if an MD's task is offloaded to a relatively far BS, data transmission may be drastically impaired due to excessive path-loss and low signal to interference plus noise ratio (SINR), which adversely affects communication resource utilization. Thus, the main challenges in MEC resource allocation can be divided into computation offloading decision (whether to offload or not), user association/offloading node selection (where to offload), subchannel assignment and transmit power control (interference management). And the optimization objectives can be divided into:

\begin{enumerate}

    \item [(1)] Energy consumption minimization, in which the energy consumption of the MDs, and/or BSs, is optimized subject to some latency constraints.

    \item [(2)] Latency minimization, where it is necessary to shorten the transmission and computation latency while satisfying some transmit power and computation resource constraints.

    \item [(3)] Joint energy consumption and latency minimization, in which the weighted sum of energy consumption and latency is minimized.

\end{enumerate}

\subsection{Related Works and Motivation}
There has been intense interest in MEC, especially in multi-user single-server scenarios, and, to a lesser extent, in homogeneous multi-server scenarios. In contrast, only a few studies have been conducted to investigate the heterogeneous servers scenario where servers are located at the SBS, MBS or cloud, and have different computing and communication capabilities, which is the focus of this paper. A summary of different MEC scenarios and objectives studied in the literature is given in Table \ref{Table1}, where it should be noted that \cite{mec20, mec21} pertain to the heterogeneous servers scenario, while the rest are for homogeneous servers. 

\setlength\arrayrulewidth{0.5pt}
\begin{table}[!ht]
\caption{Summary of MEC Scenarios and Objectives}
\centering \setlength\tabcolsep{3pt}\label{Table1}
\begin{tabular}{|c||c|c||c|c||c|c|c|}
\hline
\multirow{2}{*}{Ref.} & \multicolumn{2}{c||}{MD} & \multicolumn{2}{c||}{Edge Servers} & \multicolumn{3}{c|}{Objectives}                                 \\
\cline{2-8}
                  & Single & Multiple       & Single & Multiple                & Energy & Latency & \begin{tabular}[c]{@{}c@{}}Energy \& \\Latency\end{tabular}  \\ \hline \hline
\cite{mec20}                 &        &     \checkmark           &        &       \checkmark                  &    &         &                                                 \checkmark          \\
\cite{mec21}               &        &  \checkmark             &        &  \checkmark                      &    &  \checkmark        &                                                          \\
\cite{mec12}                 &        & \checkmark              &\checkmark        &                         &  \checkmark  &         &                                                          \\
\cite{mec13}                 &        & \checkmark               &    \checkmark    &                           &     &         &   \checkmark                                                       \\
\cite{mec14}                 &        & \checkmark              &  \checkmark      &                         &    &   \checkmark      &                                                          \\
\cite{mec15}                 &   \checkmark     &               &\checkmark        &                         & \checkmark   &         &                                                          \\
\cite{mec16}                 &  \checkmark      &               &        &       \checkmark                   &    &         &                    \checkmark                                      \\
\cite{mec17}                & \checkmark        &               &         & \checkmark                        &    &         &                                       \checkmark                    \\
\cite{mec18}                 &        &   \checkmark             &        &                   \checkmark       &   \checkmark  &         &                                                          \\
\cite{mec19}                 &        &    \checkmark           &        &      \checkmark                   &    &  \checkmark       &                                                          \\
\hline
\end{tabular}
\end{table}

On the other hand, in MEC-based systems, computational tasks offloading is potentially limited by the type of MD-BS association, and almost all the previous works consider offloading an MD's computation task to the MEC servers available at its serving BS. In 4G networks, when a typical MD accesses the network, each BS in its vicinity transmits a reference signal to it; then the MD measures the strength of this reference signal and chooses the BS with the highest biased reference signal received power (RSRP) \cite{b17}, and the chosen BS will be used for both uplink (UL) and downlink (DL) transmission. This BS access and association mode is referred to as coupled uplink/downlink access (CUDA). In homogeneous networks, CUDA is a nearly optimal access approach, since the best serving BS is same in the UL and DL \cite{a5}. However, with the deployment of more and more low-cost SBSs, the traditional homogeneous networks become heterogeneous networks (HetNets). Due to the transmit power disparity of SBSs and MBSs, the DL coverage of MBSs is usually much greater than that of SBSs. Consequently, more MDs are associated with MBSs in the DL. Offloading too many MDs to the MBSs may cause severe network congestion and latency. Furthermore, unlike BSs, MDs have similar transmit powers and transmission coverage. If the associated cell in the UL is the same as that in the DL, the link quality of the macro-cell (MCell) edge MDs will be poor, and may cause high UL transmission latency. Although cell range extension can offload more MDs to the SBSs \cite{f5}, it may impair the DL transmission performance. For balancing offloading and overcoming path-loss, it would be a better choice for some MDs to connect to a geometrically closer small-cell (SCell) in the UL rather than the same BS in the DL, so as to reduce the transmission latency. This architecture is called downlink/uplink decoupled (DUDe) access \cite{b1}. Moreover, since different types of BSs have different computation capabilities, only consider communication capability when choosing the serving BS is not enough,  taking the computation resource into consideration is also necessary. This is also a deficiency of the existing research.

\begin{figure}[htbp]
\centerline{\includegraphics[width=7cm]{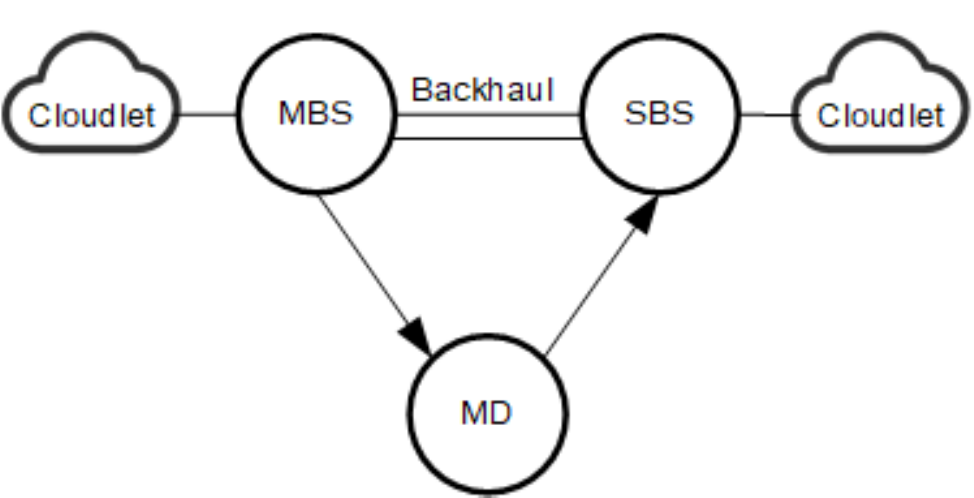}}
\caption{A DUDe MEC model.}
\label{Fig1}
\end{figure}

A typical DUDe MEC model is shown in Fig. \ref{Fig1}. According to DUDe access, an MD's UL and DL serving BSs may not necessarily be the same. Accordingly, the MD's tasks can be executed at the MEC server/cloudlet available at the UL and/or DL serving BS. In such a case, the BSs should be able to communicate with each other, which is possible for 4G and beyond cellular networks, since the BSs are inter-connected via traditional backhaul connections (e.g. via the X2 interface, in accordance with 3GPP LTE terminology) \cite{a5}. Despite the extra backhaul delay, the DUDe MEC scheme is also capable of providing a fairly lower offloading latency as compared to the conventional offloading scheme with coupled association \cite{9481951}, which highlights the need to study optimal, or near-optimal, resource allocation algorithms for DUDe-MEC systems. To the best of our knowledge, this article is the only one that applied DUDe to MEC-based networks. However, the authors did not propose any offloading or resource allocation algorithms, and the UL serving BS is chosen only based on the BS-MD distance.

\subsection{Main Contributions}
The main contributions of this paper can be summarized as follows:

\begin{itemize}

    \item A DUDe-native joint BS association and subchannel allocation algorithm is proposed. Unlike most studies that consider BS association and subchannel allocation sequentially and separately, this work simultaneously and jointly performs BS association and subchannel allocation by utilizing the student-project allocation (SPA) matching approach \cite{mec7}. Moreover, not only the communication capability, but the computation capability as well is taken into consideration when deciding the serving BS, and the constraint that one MD must connect to the same BS in the UL and DL is broken.

    \item In addition to the BS association and subchannel allocation, an optimal power allocation algorithm is devised to minimize the sum-latency of the network MDs. The formulated problem is a sum-of-ratios problem, which is non-convex and NP-hard. To efficiently and optimally solve such a problem, the proposed algorithm tightly approximates it as a convex optimization problem, and solves it successively until convergence to the global optimal allocation.

    \item To demonstrate the effectiveness of the proposed designs, a performance comparison with coupled and decoupled benchmark schemes is provided. 

\end{itemize}
The results show that the latency of the proposed DUDe access and power control scheme is much lower than that of the CUDA scheme by up to 60\%, and its EE and data rate are also better by up to 100\%.
\subsection{Organization}
The rest of the paper is organized as follows. Section \ref{sys_model} presents the system model adopted in this paper. The proposed DUDe joint BS association and subchannel allocation scheme is presented in Section \ref{jsaca_spa}, where both the communication and computation capacity of the different types of BSs are considered. Section \ref{power_allocation} presents the proposed optimal power allocation scheme to minimize the network sum-latency while section \ref{Evaluation} evaluates the performance of the proposed scheme and compares it with coupled benchmarks. Finally, Section \ref{conclusion} gives a summary and critique of the proposed algorithms as well as the findings.

\section{System Model}\label{sys_model}
\subsection{Network Model}
In this work, a two-tier orthogonal frequency-division multiple access (OFDMA) HetNet---composed of one MBS and several SBSs---is considered, and the Rayleigh fading channel model is adopted. The positions of the MBS and the SBSs are uniformly distributed, where the density of the SBSs is $\lambda_{s}$. The MDs' locations also follow a uniform distribution with a density of $\lambda_u$. Let $\mathcal{K} = \{1,\ldots,k,\ldots,K\}$ and $\mathcal{N} = \{1,2,...,N\}$ be the index sets of the MDs and subchannels, respectively. Also, let $\mathcal{M} = \{0,1,\ldots, M\}$ denote the set of $M + 1$ BSs, where BS $m = 0$ corresponds to the MBS, and the rest (i.e. $m = 1,\ldots, M$) are SBSs. Both tiers operate on the same frequency band, and are based on frequency division duplexing (FDD). Moreover, an MD can only occupy one subchannel, and MDs in different cells can reuse the same subchannel. As such, there is inter-cell interference, but no intra-cell interference.

\subsection{Transmission Model}
The transmit powers of all SBSs are assumed to be identical, but the transmit power of the MBS is greater than the SBSs. In the DL, the maximum transmit power is equally allocated over the whole bandwidth; while in the UL, the initial transmit power of MD $k$ is set according to the fractional power control (FPC), which is applied in 4G and 5G cellular networks \cite{3gpp.36.213, 3gpp.38.213}, as
\begin{eqnarray}\label{1}
    P_k^{ul} = \min\{P_{\max}, w PL + P_0\},
\end{eqnarray}where $P_{max}$ is the maximum transmit power of each MD, $P_0$ is the target received power, and $w\in \{0,0.4, 0.5,0.6,0.7,0.8,0.9,1\}$ is the compensation factor for path-loss $PL$. The path-loss $PL(d)$ can be modeled as
\begin{eqnarray}\label{2}
    PL(d) = 20 \log \left(\frac{4 \pi d_0 f}{c} \right)+10{\phi}\log\left(\frac{d}{d_0}\right)+\chi,
\end{eqnarray}where $d_0$ denotes the close-in reference distance, $f$ is the operating frequency, $c$ is the speed of light, $d$ represents the MD-BS distance, $\phi$ is the path-loss exponent, and $\chi$ is the log-normal shadowing.

\subsection{MEC Model}
In this work, full offloading is adopted, while assuming that there is only one computation task to be processed for each MD during a computation offloading period. As the UL serving BS may be different from the DL serving BS, the tasks can be executed at the cloudlet attached to the UL serving BS or the DL serving BS or both. However, for simplicity, it is assumed that all tasks are executed at the UL cloudlets in this paper. The MBS's cloudlet is assumed to have higher computation capacity than the SBSs' cloudlets, which are denoted $F^M$ and $F^S$ (in CPU cycles per second), respectively\footnote{Assume each cloudlet’s computational resources are equally shared among all tasks when two or more tasks are offloaded to the same cloudlet. Such assumption is widely applied, e.g. \cite{mec2} and \cite{mec20}.}. Also, each SBS is connected to the MBS via a backhaul link with finite capacity\footnote{Interference-free wireless backhaul links are assumed, and the backhaul links will not cause interference to MD-BS links.} $C^{bh}$. Let the computational task of each MD $k\in\mathcal{K}=\{1,2,...,K\}$ be defined as a tuple $\mathcal{T}_k \triangleq(B^C_k, B^I_k, B^O_k)$, where $B^C_k$ is the required number of CPU cycles to complete MD $k'$s task, $B^I_k$ is the number of bits of the offloaded task, and $B^O_k$ is the number of output bits representing the result of the task execution. For simplicity, the output bits and the computation data size are assumed to be proportional to the input task size. Particularly, let $B_k^0 = \alpha_k B_k^I$, where $\alpha_k$ is the proportion of output to input bits \cite{mec2}. Moreover, $B_k^C = \beta_k B_k^I$, where $\beta_k$ is the number of CPU cycles per bit, and depends on the type of executed task. To offload a task, MD $k$ first transmits the $B^I_k$ input bits to its UL serving BS, and the cloudlet of which  executes the $B_k^C$ CPU cycles. Finally, the $B^O_k$ output bits are transmitted back to the MD\footnote{Note that if the UL and DL serving BSs are different, the $B_k^O$ output bits are migrated to the DL serving BS via the backhaul link.}.

\subsection{Latency Model}
The total offloading latency for a typical MD $k$ is given by
\begin{equation}\label{3}
    L_k = L_k^{ul} + L_k^{exe} + L_k^{bh} + L_k^{dl},
\end{equation}where $L^{exe}_k$ is the time delay for the cloudlet to execute MD $k$'s task, $L_k^{ul}$ is the UL transmission latency, and $L_k^{dl}$ is the DL transmission latency\footnote{There is a waiting time before a cloudlet begins to execute a task, but most works assume it to be zero \cite{mec2, mec3}.}. Moreover, in a DUDe scenario, the computed results are migrated from the UL serving BS to the DL serving BS, and the additional migration cost is incorporated into the backhaul latency. There are two cases for the backhaul latency $L_k^{bh}$, which are 
\begin{itemize}
    \item UL serving BS $\neq$ DL serving BS: $L_k^{bh}=B_k^O/C^{bh}$;
    \item UL serving BS = DL serving BS: $L_k^{bh}=0$.
\end{itemize}

\subsection{Communication Model}
 Let $\chi_{k,m}$ be a binary decision variable, defined as
\begin{equation}\label{4}
    \begin{split}
        \chi_{k,m}^{l} = \\ & \hspace{-10mm} \begin{cases}1, & \text{if MD $k$ is with BS $m$ in $l$ direction}, \\ 0, & \text{otherwise}, \end{cases}
    \end{split}
\end{equation}where $l \in \{ul, dl\}$. Moreover, it should be noted that each MD $k \in \mathcal{K}$ can be associated with one BS in each link direction (i.e. $\sum_{m \in \mathcal{M}} \chi_{k,m}^{l} = 1$). Also, each MD is assigned one subchannel in the UL and DL directions. Thus, let $\lambda_{k,n}$ be defined as
\begin{equation}\label{5}
    \begin{split}
        \lambda_{k,n}^{l} = \\ & \hspace{-10mm} \begin{cases}1, & \text{if subchannel $n$ belongs to MD $k$ in $l$ direction}, \\ 0, & \text{otherwise}. \end{cases}
    \end{split}
\end{equation}Thus, the data rate of MD $k \in \mathcal{K}$ in the UL is
\begin{eqnarray}\label{6}
    R^{ul}_{k} = B\sum_{m \in \mathcal{M}}\sum_{n \in \mathcal{N}} \chi_{k,m}^{ul}\lambda_{k,n}^{ul} \log_2\left(1+\frac{P^{ul}_{k,m,n}h_{k,m,n}^{ul}}{I_{k,m,n}^{ul}+\sigma^2}\right),
\end{eqnarray}where $B$ is the RB bandwidth, and $\sigma^2 = N_0 B$ is the variance of the additive white Gaussian noise (AWGN), with $N_0$ being the noise spectral density. Moreover, $P^{ul}_{k,m,n}$ is the transmit power of MD $k$ when associated with BS $m$ over subchannel $n$, and $h_{k,m,n}^{ul}$ is the corresponding channel gain in the UL direction. The interference $I_{k,m,n}^{ul}$ received at BS $m$ on RB $n$ for MD $k$ is written as
\begin{eqnarray}\label{7}
    I_{k,m,n}^{ul} = \sum_{j\in \mathcal{K}, j \neq k} \lambda_{j,n}^{ul}P_{j,m,n}^{ul}h_{j,m,n}^{ul}.
\end{eqnarray}Similarly, the data rate at MD $k$ in the DL is expressed as
\begin{eqnarray}\label{8}
    R^{dl}_{k} = B\sum_{m \in \mathcal{M}}\sum_{n \in \mathcal{N}}\chi_{k,m}^{dl}\lambda_{k,n}^{dl} \log_2\left(1 + \frac{P_{k,m,n}^{dl}h_{k,m,n}^{dl}}{I_{k,m,n}^{dl}+\sigma^2}\right),
\end{eqnarray}where $P^{dl}_{k,m,n}$ is the transmit power of BS $m$ to MD $k$ over subchannel $n$, and $h_{m,k,n}^{dl}$ is the corresponding channel gain. The interference received at MD $k$ on RB $n$ is given by
\begin{eqnarray}\label{9}
    I_{m,k,n}^{dl} = \sum_{i\in \mathcal{M}, i \neq m} \lambda_{k,i}^{dl}P_{i,k,n}^{dl}h_{i,k,n}^{dl}, k\in\mathcal{K}, n\in\mathcal{N}.
\end{eqnarray}Notably, MD $k$'s UL and DL data rates $R_k^{ul}$ and $R_k^{dl}$ are related to MD $k$'s serving BS, transmit power, and the interference level over the assigned subchannel in each link direction.

The latency (in seconds) necessary to complete the UL transmission for MD $k$ is defined as
\begin{equation}\label{10}
    L_k^{ul} = \frac{B^I_k}{R_k^{ul}},
\end{equation}while the DL transmission time is
\begin{equation}\label{11}
    L_k^{dl} = \frac{B^O_k}{R_k^{dl}}.
\end{equation}The backhaul delay is related to the data size \cite{mec3}, and thus, the backhaul latency is determined as
\begin{equation}\label{12}
    L_k^{bh} = \frac{B_k^{bh}}{C^{bh}},
\end{equation}where $B^{bh}_k$ is MD $k$'s backhaul data size, defined as
\begin{equation}\label{13}
    B^{bh}_k = \begin{cases} B_k^O, & \text{decoupled access}, \\ 0, & \text{coupled access}. \end{cases}
\end{equation}The execution latency for a typical MD $k$ is
\begin{equation}\label{14}
    L_k^{exe} = \frac {B^C_k}{F_k},
\end{equation}where $F_k=\bar{F}_k/\bar{K}$, with $\bar{F}_k$ being the computation capacity (in CPU cycles/second) of the cloudlet attached to MD $k$'s UL serving BS (i.e. $\bar{F}_k = F^M$ for MBS, and $\bar{F}_k = F^S$ for SBS), and $\bar{K}$ is the number of MDs associated with that BS. In this work, a full-buffer model is assumed, and the number of MDs in a cell is constant \cite{mec20, mec2}. Such a model has been widely adopted in many OFDMA studies due to its simplicity \cite{ameigeiras2012traffic}. This reflects the trend that the more tasks/MDs are offloaded to a serving BS, the less computational resource each MD can get from that BS. Also, low mobility MDs are assumed in this paper, and hence the serving BSs remain unchanged over short time intervals \cite{mach2017mobile}. In fact, in current networks the mobility of an MD is enabled by a handover process, and when the MD moves from one place to another, it switches both the UL and DL traffic to a new BS according to the configured CUDA scheme. The number of tasks in each server (or MDs associated with a BS) will change continually, but in the simulations, most existing works assume the number of MDs/tasks is fixed over short time periods \cite{ning2018cooperative, cui2018joint, chen2015efficient}. Lastly, the total latency of MD $k$ is calculated as
\begin{equation}\label{15}
    L_k = \frac{B^I_k}{R_k^{ul}} + \frac{B^C_k}{F_k} + \frac{B_k^{bh}}{C^{bh}} + \frac{B^O_k}{R_k^{dl}}.
\end{equation}

\subsection{Problem Formulation}
In this work, the cell association, subchannel allocation and transmit power control are considered in the UL direction as DUDe can only improve the UL performance. As for DL, biased RSRP cell association is along with a greedy subchannel allocation and equal power allocation over all subchannels, with all BSs transmitting at their allowed maximum power.

Our objective function is to minimize the sum-latency, which is defined as $Z(\mathbf{X, \Lambda, \mathbf{P}}) \triangleq \Sigma_{k\in \mathcal{K}} L_k$. Specifically, $\mathbf{X} = \{\chi_{k,m}^{ul}\}$, $\mathbf{\Lambda} = \{\lambda_{k,n}^{ul}\}$, and $\mathbf{P} = \{P_{k,m,n}^{ul}\}$. Moreover, the constraints are:

\begin{itemize}

    \item Transmit power of each MD does not exceed the maximum transmit power $P_{max}$.

    \item Each MD can associate with only one BS in each link direction.

    \item Each MD-BS link can be assigned one subchannel.

\end{itemize}Therefore, the formulated problem can be expressed as

\begin{equation*}\label{16}
\hspace{-67mm}\underline{\textbf{Q1:}} \tag{16}
\end{equation*}\vspace{-0.225in}\setcounter{equation}{15}
\begin{mini!}[2]
  {\mathbf{X, \Lambda, P}}{Z(\mathbf{X, \Lambda, P}) \label{16a}}{}{}
  \addConstraint{\hspace{-2.5mm}\sum_{m \in \mathcal{M}} \chi_{k,m}}{=1,\,}{\forall k \in \mathcal{K}\label{16b}}
  \addConstraint{\hspace{-2mm}\sum_{n \in \mathcal{N}} \lambda_{k,n}}{=1,\,}{\forall k \in \mathcal{K}\label{16c}}
  \addConstraint{\hspace{-1mm}0\leq P_{k,m,n}^{ul}}{\leq P_{max},\,}{\forall k \in \mathcal{K}, \forall m \in \mathcal{M}, \forall n \in \mathcal{N}.\label{16d}}
\end{mini!}Problem \textbf{Q1} is non-convex and NP-hard. To efficiently solve it we split into two sub problems: (1) joint cell association and subchannel allocation (JCASA), and (2) power control (PC), where the FPC scheme is applied to allocate the initial UL transmit power $\mathbf{P^{(0)}}$. Then, the JCASA problem can be obtained as
\begin{equation*}\label{17}
\hspace{-67mm}\underline{\textbf{Q2:}} \tag{17}
\end{equation*}\vspace{-0.225in}\setcounter{equation}{16}
\begin{mini!}[2]
  {\mathbf{X, \Lambda}}{Z\left(\mathbf{X, \Lambda, P^{(0)}}\right) \label{17a}}{}{}
  \addConstraint{(\ref{16b}), (\ref{16c}),}{}{}
\end{mini!}which is a combinatorial problem. After obtaining the cell association and subchannel allocation solutions $\mathbf{X^*}$, and $\mathbf{\Lambda^*}$---by solving problem \textbf{Q2}---the PC problem can be written as
\begin{equation*}\label{18}
\hspace{-67mm}\underline{\textbf{Q3:}} \tag{18}
\end{equation*}\vspace{-0.225in}\setcounter{equation}{17}
\begin{mini!}[2]
  {\mathbf{P}}{Z\left(\mathbf{X^*, \Lambda^*, P}\right) \label{18a}}{}{}
  \addConstraint{(\ref{16d}),}{}{}
\end{mini!}which is a non-convex and non-linear programming problem.

In \textbf{Section \ref{jsaca_spa}}, the JCASA problem (i.e. problem \textbf{Q2}) is modeled and solved via the SPA matching approach. Thereafter, the sum-latency-minimizing power allocation algorithm for solving problem \textbf{Q3} is devised in \textbf{Section \ref{power_allocation}}.

\section{JCASA Based on SPA}\label{jsaca_spa}
\subsection{Student-Project Allocation Matching Model}
Different from the Munkres algorithm and many other matching algorithms where only two parties exist \cite{h11}, there are three parties in SPA algorithm. In the SPA matching problem, the \textbf{students} have preferences over the \textbf{projects}, while the \textbf{lecturers} have preferences over the students \cite{mec7}. In this paper, we consider the MDs $\mathcal{U} = \{u_1, \ldots, u_k, \ldots, u_K\}$ as students, the subchannels $\mathcal{C} = \{c_1, \ldots, c_{N \times (M+1)}\}$ as projects\footnote{The $N$ subchannels are assumed to be offered by the $M+1$ BSs as unique subchannels. In turn, the $N$ subchannels are replicated at each BS, such that a subchannel can be assigned to users associated with different base-stations. Thus, there are $N \times (M+1)$ subchannels in total.}, and the BSs $\mathcal{S} = \{s_0,s_1,\ldots, s_m, \ldots, s_M\}$ as the lecturers. 

Each MD $u_k$ ranks the subchannels by preference according to the SINR, i.e. if MD $u_k$ prefers subchannel $c_i$ to $c_j$, then this implies that the SINR on subchannel $c_i$---based on the allocated power---is higher than that on $c_j$. The top $N\times M_k$ subchannels with the highest SINR forms an acceptable subchannel set $\mathcal{C}_k \subseteq \mathcal{C}$, where $M_k$ is an adaptive parameter that decides if the subchannels offered by the nearest $M_k$ BSs are acceptable. Then, each BS $s_m$ ranks the set of MDs that find a project offered by it acceptable, and forms a list denoted by ${\mathcal{U}_m} \subseteq \mathcal{U}$, which consists of the set of MDs that have a suitable task size. Specifically, since the MBS has higher computation capability, it prefers the MDs that demand more computing resources, which helps reduce the network computation latency. If an MBS prefers MD $u_i$ to $u_j$, then the number of CPU cycles needed to process MD $u_i$'s task is more than that of MD $u_j$. On the other hand, if an SBS prefers MD $u_i$ to $u_j$, then the number of CPU cycles needed to process MD $u_i$'s task is less than that of MD $u_j$. For each subchannel $c_n$, let $\mathcal{U}_m^n$ denote its preference list, which is obtained from $\mathcal{U}_m$ by deleting those MDs which do not find $c_n$ acceptable. 

Each BS $s_m$ is assumed to have a capacity constraint $\Omega_m$, indicating the maximum number of MDs that it can serve at the same time, which should not exceed the total number of subchannels $N$ (i.e. $1 \leq \Omega_m \leq N$, $\forall s_m \in \mathcal{S}$). In practice, $\Omega_m$ is set as $\nu_m K/(M+1)$, which represents the average number of MDs per BS. Moreover, $\frac{M+1}{K} \leq \nu_m \leq \frac{M+1}{K}N$
is set to a high value if a BS $s_m$ is rich with computation resources, and vice versa. Also, each subchannel has a capacity constraint of 1, indicating each subchannel can be occupied by one MD when associated with a BS.
~\\
\begin{defn}[Assignment]
An assignment $\mathcal{M}$ is a subset of $\mathcal{U}\times\mathcal{C}$ such that:

\begin{enumerate}

\item [(1)] $(u_k,c_{n})\in \mathcal{M}$ (i.e. MD $u_k$ finds subchannel $c_{n}$ acceptable.

\item [(2)] $|{(u_k,c_{n})\in \mathcal{M}: c_{n} \in \mathcal{C}}|\leq 1$ (i.e. MD $u_k$ is assigned at most one subchannel).

\end{enumerate}

\end{defn}If $(u_k,c_{n})\in \mathcal{M}$, where BS $s_m$ offers subchannel $c_{n}$, then MD $u_k$ is assigned to subchannel $c_{n}$, and associated with BS $s_m$. For any MD $u_k$, $\mathcal{M}(u_k)$ refers to the subchannel it is assigned to. For any subchannel $c_{n}$, $\mathcal{M}(c_{n})$ refers the MD assigned to it. For any BS $s_m$, $\mathcal{M}(s_m)$ denotes the set of MDs associated with BS $s_m$.
~\\
\begin{defn}[Matching]
    A matching $\mathcal{M}$ is an assignment, such that $|\mathcal{M}(u_k)|\leq 1$, $|\mathcal{M}(c_{n})|\leq 1$ and $|\mathcal{M}(s_m)|\leq \Omega_m$, $\forall u_k \in \mathcal{U}$, $\forall c_{n} \in \mathcal{C}_m$, and $\forall s_m \in \mathcal{S}$.
\end{defn}

Consequently, under matching $\mathcal{M}$, each MD $u_k$ is assigned to at most one subchannel, no subchannel $c_{n} \in \mathcal{C}$ is assigned to more than one MD, and no BS $s_m$ is assigned to more than $\Omega_m$ MDs.
~\\
\begin{defn}[Subscription]
    A subchannel $c_{n} \in \mathcal{C}_m$ is said to be under-subscribed, full, or over-subscribed if $|\mathcal{M}(c_{n})| < 1$, $|\mathcal{M}(c_{n})| = 1$, or $|\mathcal{M}(c_{n})|> 1$, respectively. Similarly, a BS $s_m \in \mathcal{S}$ is considered to be under-subscribed, full, or over-subscribed if $|\mathcal{M}(s_m)| < \Omega_m$, $|\mathcal{M}(s_m)| = \Omega_m$, or $|\mathcal{M}(s_m)| > \Omega_m$, respectively.
\end{defn}
~\\
\begin{defn}[Blocking]
    The pair $|(u_k,c_n)\in(\mathcal{U}\times\mathcal{C})\backslash \mathcal{M}$ is said to block a matching $\mathcal{M}$ if: 

(a) $c_n\in \mathcal{C}_k$ (i.e. $u_k$ finds $c_n$ acceptable).

(b) Either $u_k$ is unassigned in $\mathcal{M}$, or $u_k$ prefers $c_n$ to ${M}(u_k)$.

(c) Either

(c1) $c_n$ is under-subscribed and $s_m$ is under-subscribed, or

(c2) $c_n$ is under-subscribed, $s_m$ is full, and either $u_k \in \mathcal{M}(s_m)$ or $s_m$ prefers $u_k$ to the worst MD in $\mathcal{M}(s_m)$, or

(c3) $c_n$ is full and $s_m$ prefers $u_k$  to the worst MD in $\mathcal{M}(c_n)$, where $s_m$ is the BS who provides $c_n$.
\end{defn}
~\\
\begin{defn}[Stable matching]
    A matching $\mathcal{M}$ is considered stable if $\mathcal{M}$ contains no blocking pairs.
\end{defn}
~\\
\hspace*{0.4cm}In contrast to most previous works, which consider BS association and subchannel allocation sequentially and separately, the SPA algorithm performs such tasks simultaneously and jointly. Initially, all MDs are free, and all subchannels and BSs are unsubscribed. As long as there is an MD, $u_k$, that is free and with a non-empty preference list, it can apply to the first subchannel $c_n$ on $\mathcal{C}_k$. Let $s_m$ be the BS that offers $c_n$. Immediately, $u_k$ becomes provisionally assigned to $c_n$ (and to $s_m$). If $c_n$ is over-subscribed, then $s_m$ rejects the worst MD $u_r$ assigned to $c_n$ and the pair $(u_r, c_n)$ will be deleted. Similarly, if BS $s_m$ is over-subscribed, then $s_m$ rejects its worst assigned MD $u_r$ and the pair $(u_r, c_t)$ will be deleted from $\mathcal{M}$, where $c_t$ was the subchannel assigned to $u_r$. On the other hand, if $c_n$ is full and $u_r$ is the worst MD assigned to $c_n$, then delete $(u_t, c_n)$, where $u_t$ is each successor of $u_r$ on $\mathcal{U}_m^n$. Similarly, if $s_m$ is full and $u_r$ is the worst MD assigned to $s_m$, then delete $(u_t, c_v)$, where $u_t$ is each successor of $u_r$ on $\mathcal{U}_m$ and $c_v$ is each subchannel offered by $s_m$ that $u_t$ finds acceptable. The SPA algorithm is described in \textbf{Algorithm \ref{algorithm_1}}. 

\begin{rem}It should be noted that the accept and/or reject processes occur during the execution of \textbf{Algorithm \ref{algorithm_1}} at a network centralized node (as in the case of a centralized radio access network (C-RAN))\footnote{The convergence of the MEC and C-RAN technology is a new trend in 5G and beyond, with operators making cost savings by deploying C-RAN and MEC infrastructures together}, and not during the network operation\footnote{The centralized node has the preference lists of all MDs and BSs in an area, and can make the BS association and subchannel allocation decisions.}. That is, the network operation takes place after the centralized node gives the BS association and subchannel allocation solutions. Therefore, it will not cause call drops or failed handovers, and there is no data exchange between MDs and BSs during the execution of the SPA algorithm. Additionally, once a subchannel is occupied during the network operation, it will not be re-allocated to other MDs.
\end{rem}
\begin{algorithm}
\caption{SPA based JCASA}\label{algorithm_1}
\begin{algorithmic}[1]
\STATE  \textbf{Input:} Preference lists $\mathcal{U}_m$, $\mathcal{U}_m^n$ and $\mathcal{C}_k$, for $\forall s_m\in \mathcal{S}$, $\forall c_n\in \mathcal{C}$ and $\forall u_k\in \mathcal{U}$.
\STATE  \textbf{Initialization:} All MDs are free, and all subchannels and BSs are unsubscribed.
\WHILE { (Some MD $u_k$ is free and has a non-empty list $\mathcal{C}_k$)}
\STATE  $c_n$ = first subchannel on $\mathcal{C}_k$;
\STATE  $s_m$ = the BS that offers $c_n$;
\STATE  provisionally assign subchannel $c_n$ and BS $s_m$ to MD $u_k$;
\IF { ($c_n$ is over-subscribed)}
\STATE  $u_r$ = worst MD assigned to subchannel $c_n$;
\STATE  $s_m$ = the BS that offers subchannel $c_n$;
\STATE  $s_m$ rejects $u_r$ and break assignment $(u_r,c_n)$;
\ELSE
\IF{{ ($s_m$ is over-subscribed)}}
\STATE  $u_r$ = worst MD assigned to BS $s_m$;
\STATE  $c_t$ = the subchannel assigned to MD $u_r$;
\STATE  $s_m$ rejects $u_r$ and break assignment $(u_r,c_t)$;
\ENDIF
\ENDIF
\ENDWHILE
\IF {($c_n$ is full)}
\STATE  $u_r$ = worst MD assigned to subchannel $c_n$;
\STATE  $u_l$ = each of the successor of $u_r$ on $\mathcal{U}_m^n$;
\STATE  Break assignment $(u_l,c_n)$;
\ENDIF
\IF {($s_m$ is full)}
\STATE  $u_r$ = worst MD assigned to BS $s_m$;
\STATE  $u_l$ = each successor of $u_r$ on $\mathcal{U}_m$;
\STATE  $c_v$ = each subchannel offered by $s_m$ that $u_t$ finds acceptable;
\STATE  Break assignment $(u_l,c_v)$;
\ENDIF
\STATE  \textbf{Output:} Stable matching $(u_k,c_n)\in \mathcal{U}\times\mathcal{C}$.
\end{algorithmic}
\end{algorithm}

\subsection{Swap Matching}
SPA-based JCASA is with externalities (also known as peer effects) \cite{mec8}. The preference lists of the MDs and subchannels are initially constructed without considering the inter-cell interference over each subchannel, since no BS association or subchannel allocation initially exists. Thus, although \textbf{Algorithm \ref{algorithm_1}} yields a stable matching, as soon as the subchannels are assigned to the MDs, the appearance of inter-cell interference terms may not necessarily preserve the stability of the assignment. For this reason, swap matching is adopted among the MDs associated with the same BS.
~\\
\begin{defn}[Swap-Blocking Pair]
    Two subchannels $c_{n}$ (assigned to MD $u_k$) and $c_{n'}$ (assigned to MD $u_{k'}$, or not occupied)---for $n' \neq n$ and $k' \neq k$---offered by BS $s_m$ form a swap-block pair if:

    \begin{enumerate}

        \item [(a)] MD $u_k$ can get higher data rate on subchannel $c_{n'}$,

        \item [(b)] subchannel $c_{n'}$ is assigned to MD $u_{k'}$, but MD $u_{k'}$ can get higher data rate on subchannel $c_{n}$, and the sum UL transmission latency of MDs $u_k$ and $u_{k'}$ is lower, and

        \item [(c)] sum-latency of all MDs transmitting over subchannel $c_{n}$ and $c_{n'}$ is lower after the swap\footnote{Note that swapping users within one BS may trigger swaps for users in other BS. This is due to the frequency re-use.}.

    \end{enumerate}

\end{defn}

Given a matching $\mathcal{M}$, when a swap-blocking pair ($c_{n}$, $c_{n'}$) exists, MDs $u_k$ and $u_{k'}$ swap their subchannels, while keeping the other MD and subchannel assignments unchanged. In turn, the updated matching is obtained as

\begin{align*}&\hspace {-2pc}\overline {\mathcal {M}} = \left \{\mathcal {M} \setminus \left \{{\left ({u_{k}, \mathcal {M}\left ({u_{k}}\right) }\right), \left ({u_{r},\mathcal {M}\left ({u_{r} }\right) }\right) }\right \} \right \} \\&\qquad  \cup \left \{{\left ({u_{k}, \mathcal {M}\left ({u_{r}}\right) }\right), \left ({u_{r}, \mathcal {M}\left ({u_{k} }\right) }\right) }\right \}, \tag{19}\end{align*}
On the other hand, if $c_{n',m}$ is not occupied, then
\begin{align*}&\hspace {-2pc}\overline {\mathcal {M}} = \left \{{\mathcal {M} \setminus \left \{{\left ({u_{k}, \mathcal {M}\left ({u_{k}}\right) }\right)}\right \} }\right \} \cup \left \{{\left ({u_{k}, \mathcal {M}\left ({u_{r}}\right) }\right) }\right \}. \tag{20}\end{align*}
~\
\begin{defn}[Two-Sided Exchange-Stability]
    A matching $\overline{\mathcal{M}}$ is said to be two-sided exchange-stable if it does not contain any swap-blocking pairs \cite{mec9}.
\end{defn}

The swap matching algorithm is described as follows. Each MD searches if there is a subchannel that can provide a higher data rate. If there is such a subchannel, check if it can form a swap-blocking pair. If a swap-blocking pair is found, then a swap-operation is performed, and the matching is updated. This process is given in \textbf{Algorithm \ref{algorithm_2}}. 

\begin{algorithm}
\caption{Swap Matching for MDs over the same BS}\label{algorithm_2}
\begin{algorithmic}[1]
\STATE \textbf{Input:} Matching $\mathcal {M}$
\FOR {Each MD assign to $c_{n}$}
\IF {there is a subchannel $c_{n'}$ that can provide higher data rate}
\IF {$(c_{n}, c_{n'})$ is a swap-blocking pair}
\STATE Perform a swap-operation;
\STATE Update matching $\overline {\mathcal {M}}$;
\ENDIF
\ENDIF
\ENDFOR
\STATE  \textbf{Output:} Updated matching $\overline {\mathcal {M}}$.
\end{algorithmic}
\end{algorithm}

\subsection{Properties}
The SPA algorithm converges within a polynomial-time complexity of $\mathcal{O}(|\mathcal{U}|\times|\mathcal{C}|)$, where $|\mathcal{U}|=K$ and $|\mathcal{C}|=N\times (M+1)$ are the number of MDs and subchannels, respectively. The stable matching resulting from the SPA algorithm is optimal with respect to each assigned MD \cite{mec7}. This is because each user is assigned to its most preferred subchannel available on its preference list, and no stable pair is deleted during the execution of the SPA algorithm \cite{mec7}. In turn, each user is simultaneously assigned to the best channel it can get in any stable matching.

As for \textbf{Algorithm \ref{algorithm_2}}, it is guaranteed to converge to a matching $\overline{\mathcal{M}}$ in a finite number of iterations, since the number of subchannel pairs that can be swapped for each MD pair is finite. Note that there are $\binom {N}{2} = \frac{1}{2}\left({N^{2}-N}\right)$ subchannel pairs. Thus, the worst-case complexity is $\mathcal{O}\left({N^{2}}\right)$ per user pair.


\section{Power Allocation}\label{power_allocation}
Since this work is focussed on minimizing the sum-latency, and does not optimize the DL latency, problem \textbf{Q3} can be simplified as

\begin{equation*}\label{21}
\hspace{-67mm}\underline{\textbf{Q4:}} \tag{21}
\end{equation*}\vspace{-0.225in}\setcounter{equation}{20}
\begin{mini!}[2]
  {\mathbf{P}}{\sum_{k\in \mathcal{K}} \left(\frac{B^I_k}{R_k^{ul}(\mathbf{P})}+\frac{B^C_k}{F_k}+\frac{B_k^{bh}}{C^{bh}}\right) \label{21a}}{}{}
  \addConstraint{0\leq P_k^{ul}}{\leq P_{max},\quad}{\forall k \in \mathcal{K}\label{21b}}.
\end{mini!}For each MD, when $\mathbf{X, \Lambda}$ are fixed, the term $\frac{B^C_k}{F_k}+\frac{B_k^{bh}}{C^{bh}}$ is also fixed. Thus, \textbf{Q4} can be converted to
\begin{equation*}\label{22}
\hspace{-67mm}\underline{\textbf{Q5:}} \tag{22}
\end{equation*}\vspace{-0.225in}\setcounter{equation}{21}
\begin{mini!}[2]
  {\mathbf{P}}{\sum_{k\in \mathcal{K}} \frac{B^I_k}{R_k^{ul}(\mathbf{P})} \label{22a}}{}{}
  \addConstraint{0\leq P_k^{ul}}{\leq P_{max},\quad}{\forall k \in \mathcal{K}\label{22b}}.
\end{mini!}Problem \textbf{Q5} is non-convex and NP-hard. Alternatively, it can be transformed as
\begin{equation*}\label{23}
\hspace{-67mm}\underline{\textbf{Q6:}} \tag{23}
\end{equation*}\vspace{-0.225in}\setcounter{equation}{22}
\begin{mini!}[2]
  {\mathbf{P}, \boldsymbol{\tau}}{\sum_{k\in \mathcal{K}} \tau_k \label{23a}}{}{}
  \addConstraint{\frac{B^I_k}{R_k^{ul}(\mathbf{P})}}{\leq \tau_k,\quad}{\forall k \in \mathcal{K}\label{23b}}
  \addConstraint{0\leq P_k^{ul}}{\leq P_{max},\quad}{\forall k \in \mathcal{K}\label{23c}}.
\end{mini!}

\begin{prop}\label{prop1}
    If ($\mathbf{P}^*, \boldsymbol{\tau}^*$) is the solution of Problem \textbf{Q6}, then there exists $\boldsymbol{\lambda}^* = [\lambda_1,\lambda_2,...,\lambda_k]$ such that $\mathbf{P}^*$ satisfies the KKT conditions of the following problem upon setting $\boldsymbol{\lambda} = \boldsymbol{\lambda}^*$ and $\boldsymbol{\tau} = \boldsymbol{\tau}^*$,

\begin{equation*}\label{24}
\hspace{-64mm}\underline{\textbf{Q7:}} \tag{24}
\end{equation*}\vspace{-0.225in}\setcounter{equation}{23}
\begin{mini!}[2]
  {\mathbf{P}}{\sum_{k\in \mathcal{K}} \lambda_k(B^I_k-\tau_k{R_k^{ul}(\mathbf{P}))} \label{24a}}{}{}
  \addConstraint{0\leq P_k^{ul}}{\leq P_{max},\quad}{\forall k \in \mathcal{K}\label{24b}}.
\end{mini!}
\end{prop}

\begin{IEEEproof}
The Lagrangian of Problem \textbf{Q6} is expressed as
\begin{equation}\label{25}
    \mathcal{L}(\boldsymbol{\lambda}, \mathbf{P}, \boldsymbol{\tau}) = \sum_{k\in \mathcal{K}} \tau_k + \sum_{k\in \mathcal{K}} \lambda_k\left(B^I_k - \tau_k R_k^{ul}(\mathbf{P})\right).
\end{equation}
If ($\mathbf{P}^*, \boldsymbol{\tau}^*$) is the solution of Problem \textbf{Q6}, then there exists $\boldsymbol{\lambda}^*$ satisfying the following KKT conditions, $\forall k\in \mathcal{K}$, such that

\begin{equation}\label{26}
    \frac{\partial \mathcal{L}(\boldsymbol{\lambda}, \mathbf{P}, \boldsymbol{\tau})}{\partial \tau_k}=1-\lambda_k^*R_k^{ul}(\mathbf{P}^*) = 0,
\end{equation}
\begin{equation}\label{27}
    \frac{\partial \mathcal{L}(\boldsymbol{\lambda}, \mathbf{P}, \boldsymbol{\tau})}{\partial P_k}=-\sum_{k\in \mathcal{K}} \lambda_k^*\tau_k^* \frac{\partial R_k^{ul}(\mathbf{P}^*)}{\partial P_k} = 0,
\end{equation}
\begin{equation}\label{28}
    \lambda_k^*(B^I_k-\tau_k^* R_k^{ul^*}(\mathbf{P})) = 0,
\end{equation}
\begin{equation}\label{29}
    B^I_k-\tau_k^* R_k^{ul^*}(\mathbf{P}) \leq 0,
\end{equation}
\begin{equation}\label{30}
    \lambda_k^* \geq 0,
\end{equation}and
\begin{equation}\label{31}
{0\leq P_k^{ul}}{\leq P_{max}}.
\end{equation}

According to (\ref{26}), it can be verified that
\begin{equation}\label{32}
    \lambda_k^* = \frac{1}{R_k^{ul}(\mathbf{P}^*)} > 0.
\end{equation}Furthermore, (\ref{28}) implies that
\begin{equation}\label{33}
    \tau_k^*=\frac{B^I_k}{R_k^{ul}(\mathbf{P}^*)} > 0.
\end{equation}Lastly, since (\ref{26}), (\ref{27}) and (\ref{31}) are also the KKT conditions of Problem \textbf{Q7}, Proposition 1 is proved.
\end{IEEEproof}

\begin{figure*}[!ht]\setcounter{equation}{37}
\begin{equation}\label{38}
    \begin{split}
        R^{ul}_k(\mathbf{P}) = B \sum_{m \in \mathcal{M}}\sum_{n \in \mathcal{N}}\chi_{k,m}^{ul}\lambda_{k,n}^{ul} \log_2\left(1 + \frac{P^{ul}_{k}h_{k,m,n}^{ul}}{\sum_{j\in \mathcal{K}, j \neq k} \lambda_{j,n}^{ul}P_{j}^{ul}h_{j,m,n}+\sigma^2}\right)
\end{split}
\end{equation}\hrule\end{figure*}
\begin{figure*}\setcounter{equation}{41}
\begin{equation}\label{42}
\begin{split}
R^{ul}_k(\mathbf{P}) & \geq \overline{R}^{ul}_{k}(\mathbf{Q}) \\ & \triangleq B\left(\sum_{m \in \mathcal{M}}\sum_{n \in \mathcal{N}}\chi_{k,m}^{ul}\lambda_{k,n}^{ul} \left(\mu_{1,k}\log_2\left(\frac{2^{Q^{ul}_{k}}h_{k,m,n}^{ul}}{\sum_{j\in \mathcal{K}, j \neq k} \lambda_{j,n}^{ul}2^{Q_{j}^{ul}}h_{j,m,n}+\sigma^2}\right) + \mu_{2,k}\right) \right) \\ & = B\left(\sum_{m \in \mathcal{M}}\sum_{n \in \mathcal{N}}\chi_{k,m}^{ul}\lambda_{k,n}^{ul} \left(\mu_{1,k}Q^{ul}_{k} + \mu_{1,k}\log_2(h_{k,m,n}^{ul}) - \mu_{1,k}\log_2 \left(\sum_{j\in \mathcal{K}, j \neq k} \lambda_{j,n}^{ul}2^{Q_{j}^{ul}}h_{j,m,n}+\sigma^2\right) + \mu_{2,k}\right) \right)
\end{split}
\end{equation}\hrule \vspace{-0.05in}\end{figure*}

Based on \textbf{Proposition \ref{prop1}}, when $\boldsymbol{\lambda} = \boldsymbol{\lambda}^*$ and $\boldsymbol{\tau} = \boldsymbol{\tau}^*$, $P^*$ satisfies the following constraints \setcounter{equation}{33}
\begin{equation}\label{34}
    \lambda_k=\frac{1}{R_k^{ul}(\mathbf{P}^*)}, \quad \forall k\in \mathcal{K},
\end{equation} and
\begin{equation}\label{35}
    \tau_k=\frac{B^I_k}{R_k^{ul}(\mathbf{P}^*)}, \quad \forall k\in \mathcal{K}.
\end{equation}

In this way, Problem \textbf{Q6} is transformed into \textbf{Q7}, which can be solved in two steps. Firstly, fix $\boldsymbol{\lambda}$ and $\boldsymbol{\tau}$, and obtain $\mathbf{P}$ by solving Problem \textbf{Q7} via \textbf{Algorithm \ref{algorithm_4}}, which will be discussed shortly. Then, fix $\mathbf{P}$, and update $\boldsymbol{\lambda}$ and $\boldsymbol{\tau}$ via the modified Newton's algorithm until convergence \cite{mec24, mec10}. The process is given in \textbf{Algorithm \ref{algorithm_3}}, where
\begin{equation}\label{36}
    \rho_k(\lambda_k) = \lambda_k{R_k^{ul}(\mathbf{P}^*)}-1,\quad \forall k\in \mathcal{K},
\end{equation}and
\begin{equation}\label{37}
    \kappa_k(\tau_k) = \tau_k R_k^{ul}(\mathbf{P}^*)-B^I_k,\quad \forall k\in \mathcal{K}.
\end{equation}

\begin{algorithm}[!ht]
\caption{Optimal Power Allocation}\label{algorithm_3}
\begin{algorithmic}[1]
\STATE \textbf{Initialization:} $\mathbf{P}^{(0)}$, $t=0$, $\zeta\in(0,1)$, $\epsilon\in(0,1)$, calculate $\lambda_k$ and $\tau_k$ by (\ref{34}) and (\ref{35}), respectively.
\REPEAT
\STATE Update $\mathbf{P}^{(t+1)}$ via \textbf{Algorithm 4};
\STATE Update $\lambda_k$ and $\tau_k$ as follows
\begin{equation*}
    \lambda_k^{(t+1)}=\lambda_k^{(t)}-\frac{\zeta^{i^{(t+1)}}\rho_k\left(\lambda_k^{(t)}\right)}{R_k^{ul}(\mathbf{P}^{(t+1)})}, \quad \forall k\in \mathcal{K},
\end{equation*}and
\begin{equation*}
    \tau_k^{(t+1)}=\tau_k^{(t)}-\frac{\zeta^{i^{(t+1)}}\kappa_k\left(\tau_k^{(t)}\right)}{R_k^{ul}(\mathbf{P}^{(t+1)})}, \quad \forall k\in \mathcal{K},
\end{equation*}where $i^{(t+1)}$ is the smallest integer among $i \in \{1, 2, 3, . . .\}$ satisfying
\small
\begin{equation*}
\begin{split}
\sum_{k\in \mathcal{K}}\biggl|\rho_k\left(\lambda_k^{(t)} - \frac{\zeta^{i}\rho_k\left(\lambda_k^{(t)}\right)}{R_k^{ul}(\mathbf{P}^{(t+1)})}\right)\biggr|^2 \\ & \hspace{-42.5mm} + \sum_{k\in \mathcal{K}}\biggl|\kappa_k\left(\tau_k^{(t)} - \frac{\zeta^{i}\kappa_k\left(\tau_k^{(t)}\right)}{R_k^{ul}(\mathbf{P}^{(t+1)})}\right)\biggr|^2 \\ & \hspace{-40mm} \leq \left(1 - \epsilon\zeta^{i}\right)^2\sum_{k\in \mathcal{K}}\left(\left|\rho_k\left(\lambda_k^{(t)}\right)\right|^2+\left|\kappa_k\left(\tau_k^{(t)}\right)\right|^2\right);
\end{split}\end{equation*}

\STATE  Set $t=t+1$;
\UNTIL{the following conditions are satisfied:}
\begin{equation*}
    \lambda_k^{(t)}{R_k^{ul}\left(\mathbf{P}^{(t)}\right)} - 1 = 0, \quad \forall k \in \mathcal{K},
\end{equation*}and
\begin{equation*}
    \tau_k^{(t)} R_k^{ul}\left(\mathbf{P}^{(t)}\right) - B^I_k = 0,\quad \forall k \in \mathcal{K};
\end{equation*}
\STATE \small \textbf{Output:} Optimal $\mathbf{P}^*=\mathbf{P}^{(t)}$.
\end{algorithmic}
\end{algorithm}

In order to solve Problem \textbf{Q7}, which is non-convex since the rate function $R_k^{ul}(\mathbf{P})$ is non-convex, it is transformed into a convex problem. Note that the rate function in (\ref{6}) can be more conveniently written as (\ref{38}). Now, consider the lower-bound approximation \cite{m1} \setcounter{equation}{38}\vspace{-0.1in}

\begin{equation}\label{39}
    \log_2(1 + \gamma) \geq \mu_1 \log_2(\gamma) + \mu_2,
\end{equation}where $\gamma\geq 0$, and the bound is tight for $\gamma = \bar{\gamma}$. Moreover,
\begin{equation}\label{40}
    \mu_1 = \frac{\bar{\gamma}}{\bar{\gamma} + 1},
\end{equation}and
\begin{equation}\label{41}
    \mu_2 = \log_2(1 + \bar{\gamma}) - \mu_1 \log_2 (\bar{\gamma}).
\end{equation}By using the variable substitution $P^{ul}_k = 2^{Q^{ul}_k}$, the rate function can be lower-bounded as (\ref{42}). Hence, Problem \textbf{Q7} can be rewritten as
\begin{equation*}\label{43}
\hspace{-64mm}\underline{\textbf{Q8:}} \tag{43}
\end{equation*}\vspace{-0.225in}\setcounter{equation}{42}
\begin{mini!}[2]
  {\mathbf{Q}}{\bar{L}(\mathbf{Q})=\sum_{k\in \mathcal{K}} \lambda_k(B^I_k-\tau_k\overline{R}^{ul}_{k}(\mathbf{Q}))\label{43a}\tag{43a}}{}{}
  \addConstraint{0\leq 2^{Q_k^{ul}}}{\leq P_{max},\quad}{\forall k \in \mathcal{K}.\label{43b}\tag{43b}}
\end{mini!}

\begin{algorithm}
\caption{Solution of Problem \textbf{Q8}}\label{algorithm_4}
\begin{algorithmic}[1]
\STATE \textbf{Initialization:} Set error tolerance $\epsilon\in (0,1)$, iteration index $t = 0$, $\mu_{1,k} = 1$, $\mu_{2,k} = 0$, $\forall k \in \mathcal{K}$, select a feasible $\mathbf{Q}^{(0)}$, and calculate $\bar{L}(\mathbf{Q}^{(0)})$.
\REPEAT
\STATE Set $t = t+1$;
\STATE Update $\mu^{(t)}_{1,k}$ and $\mu^{(t)}_{2,k}$ by (\ref{40}) and (\ref{41}), $\forall k \in \mathcal{K}$;
\STATE Compute $\bar{L}(\mathbf{Q}^{(t)})$ by solving Problem \textbf{Q8};
\UNTIL $\left|\bar{L}(\mathbf{Q}^{(t)})-\bar{L}(\mathbf{Q}^{(t-1)}) \right|\leq\epsilon$
\STATE  \textbf{Output:} $\mathbf{P}^{\star}=2^{\mathbf{Q}^{(t)}}$
\end{algorithmic}
\end{algorithm}

\subsection{Properties}
For the rate function $\overline{R}^{ul}_{k}(\mathbf{Q})$ in (\ref{42}), the negative log-sum-exp term is concave in $\mathbf{Q}$ \cite{m2}. Thus, $\overline{R}^{ul}_{k}(\mathbf{Q})$ is concave and $\lambda_k\left(B^I_k-\tau_k\overline{R}^{ul}_{k}(\mathbf{Q})\right)$ is convex. More importantly, the sum of the convex functions is also convex \cite{m3}, and hence, the lower-bounded objective function $\sum_{k\in \mathcal{K}} \lambda_k\left(B^I_k-\tau_k\overline{R}^{ul}_{k}(\mathbf{Q})\right)$ is convex in $\mathbf{Q}$. Also, the constraint set of Problem \textbf{Q8} is convex. Accordingly, Problem \textbf{Q8} can be solved optimally for fixed values of $\mu_{1,k}$ and $\mu_{2,k}$ via any standard convex optimization package. By iteratively updating $\mu_{1,k}$ and $\mu_{2,k}$ via (\ref{40}) and (\ref{41}), respectively, $\mathbf{P}^{(t+1)}$ in \textbf{Algorithm \ref{algorithm_3}} is obtained via \textbf{Algorithm \ref{algorithm_4}}. Hence, the global optimal solution of problem \textbf{Q5} can be obtained.

The complexity of \textbf{Algorithm \ref{algorithm_3}} is mainly dependent on Step 3, as all the other steps are based on explicit expressions. Since a convex optimization problem is solved in each iteration in \textbf{Algorithm \ref{algorithm_4}}, it has polynomial-time complexity \cite{m3}. As for convergence, \textbf{Algorithm \ref{algorithm_3}} is guaranteed to converge in a finite number of iterations \cite{mec24}.

\subsection{Summary of Proposed Scheme}
\begin{figure}[htbp]
\centerline{\includegraphics[width=6cm]{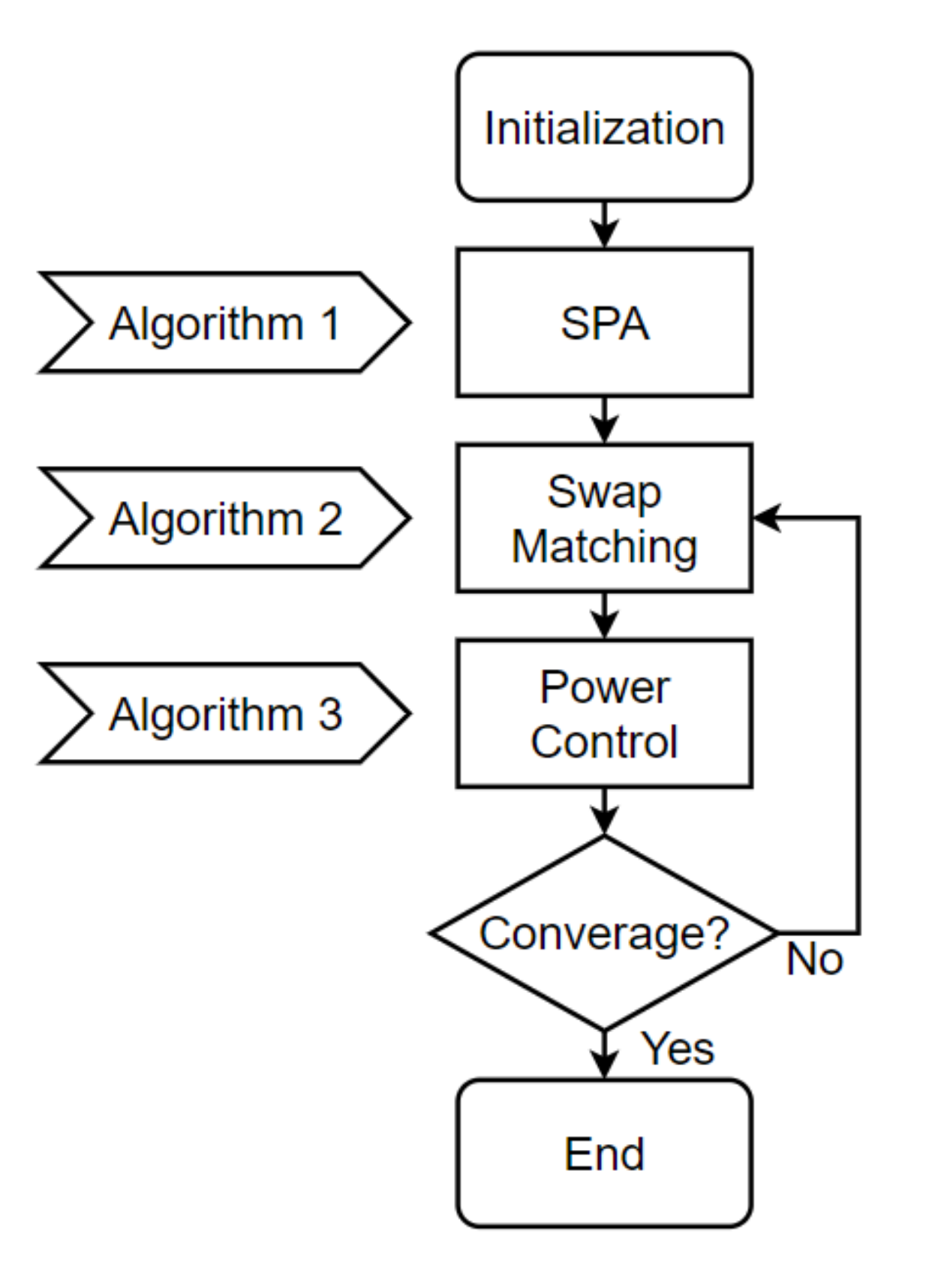}}\vspace{-0.1in}
\caption{Flow-chart of the proposed scheme.}
\label{Fig2}
\end{figure}

A flow-chart of the proposed joint BS association, subchannel allocation, and power control scheme is given in Fig. \ref{Fig2}. The proposed scheme starts by performing BS association and subchannel allocation via the SPA algorithm given in \textbf{Algorithm \ref{algorithm_1}}. After that, if at least one swap-blocking pair is found, then the corresponding MDs' subchannels are swapped via \textbf{Algorithm \ref{algorithm_2}}. Then, optimal power allocation is applied using \textbf{Algorithms \ref{algorithm_3}} and \textbf{\ref{algorithm_4}}. After the power allocation, if a swap-blocking pair is found, \textbf{Algorithm \ref{algorithm_2}} is applied again, which is followed by optimal power allocation, and so on. This is to ensure stability after power allocation until no further swap-blocking pairs can be found\footnote{Generally speaking, it has been determined that the possibility of finding a swap-blocking pair to perform swap matching is very low after the first loop.}.

\section{Performance Evaluation}\label{Evaluation}
In this section, the performance of the proposed DUDe access scheme is evaluated and compared with coupled and decoupled benchmark schemes in terms of latency and energy-efficiency (EE). The network energy-efficiency in the UL is determined as \cite{a2}
\begin{eqnarray}\label{48}
    EE = \frac{\sum_{k\in \mathcal{K}}R^{ul}_k }{\sum_{k\in \mathcal{K}}\sum_{m \in \mathcal{M}}\sum_{n \in \mathcal{N}}P_{k,m,n}^{ul}}.
\end{eqnarray}

In what follows, the proposed SPA algorithm along with swap matching (SM), and optimal power allocation (OPA) are compared to the CUDA scheme, which is based on biased RSRP for cell association, greedy (G) subchannel allocation \cite{h17}\footnote{In the greedy subchannel allocation algorithm, subchannels with high SINR are preferentially assigned to the MDs \cite{h17}.}, and FPC. Moreover, a decoupled access scheme, called Min-PL-F-FPC is also compared, which is based on minimum path-loss (min-PL) cell association \cite{b1}, greedy subchannel allocation, and FPC. In the min-PL criterion, the MDs are connected in the UL to the BS with the lowest path-loss. Particularly, a typical MD $k \in \mathcal{K}$ is associated with BS $m \in \mathcal{M}$ in the UL if
\begin{eqnarray}\label{49}
    P_{k,m}^{ul}{W_m}{PL_{k,m}}^{-1} \ge {P_{k,m'}^{ul}{W_{m'}}{PL^{-1}_{{k,m'}}}}, \quad \forall m,m' \in \mathcal{M},
\end{eqnarray}where $W$ is the UL cell bias value, which is positive and refers to expanding the coverage of the cells. Moreover, ${PL_{k,m}}$ is the UL path-loss between MD $k$ and BS $m$. Table \ref{Table2} summarizes the evaluated resource allocation schemes, while the simulation parameters are given in Table \ref{Table3}, and are obtained from related works.

\begin{table}[!ht]
\caption{Resource Allocation Schemes}
\centering \label{Table2}\setlength\tabcolsep{2.0pt}\vspace{-0.1in}
        \setlength\arrayrulewidth{1pt}
\begin{tabular}{|p{0.21\columnwidth}||p{0.19\columnwidth}|p{0.19\columnwidth}|p{0.20\columnwidth}|}
\hline
\multirow{2}{*}{\textbf{Scheme}} & \textbf{Cell-Association} & {\textbf{Subchannel Allocation}} & \multirow{2}{*}{\textbf{Power Control}}\\
\hline \hline
CUDA & Biased RSRP & Greedy & FPC \\
\hline
Min-PL-G-FPC & Min-PL & Greedy & FPC \\
\hline
SPA-FPC & \multicolumn{2}{c|}{SPA} & FPC \\
\hline
SPA-SM-FPC & \multicolumn{2}{c|}{SPA-SM} & FPC \\
\hline
SPA-SM-OPA & \multicolumn{2}{c|}{SPA-SM} & OPA \\
\hline
\end{tabular}
\end{table}

\begin{table}[!ht]
\caption{Simulation Parameters}
\centering \label{Table3}\setlength\tabcolsep{3.6pt}\vspace{-0.1in}
        \setlength\arrayrulewidth{1pt}
\begin{tabular}{|p{0.45\columnwidth}||p{0.12\columnwidth}|p{0.12\columnwidth}|p{0.17\columnwidth}|}
\hline
\textbf{Parameters} & \textbf{MD}  & \textbf{MBS} & \textbf{SBS}\\
\hline \hline
Maximum transmit power & 23 dBm  & 46 dBm & 30 dBm\\
\hline
Spatial density & 250/km$^2$ & 5/km$^2$ & 25/km$^2$\\
\hline
Spatial distribution & \multicolumn{3}{c|}{Uniform distribution}\\
\hline
Lognormal shadowing & \multicolumn{3}{c|}{$\mu=0$, $\sigma=4$ dB \cite{i1}} \\
\hline
Path-loss exponent & \multicolumn{3}{c|}{3 \cite{i1}} \\
\hline
Operating frequency & \multicolumn{3}{c|}{2 GHz}\\
\hline
UL bandwidth & \multicolumn{3}{c|}{5 MHz \cite{c2}}\\
\hline
DL bandwidth & \multicolumn{3}{c|}{5 MHz \cite{c2}}\\
\hline
Subcarrier spacing & \multicolumn{3}{c|}{15 kHz \cite{c2}} \\
\hline
Number of subchannels & \multicolumn{3}{c|}{25}\\
\hline
Noise spectral density & \multicolumn{3}{c|}{-174 dBm/Hz}\\
\hline
Target received power $P_0$ & \multicolumn{3}{c|}{-80 dBm}\\
\hline
PL compensation factor $w$ & \multicolumn{3}{c|}{0.7}\\
\hline
Computation capacity & N/A & 36 GHz & 3.6GHz \cite{mec3}\\
\hline 
Backhaul link capacity $C^{bh}$ & \multicolumn{3}{c|}{10 Mbits/s \cite{mec3}} \\
\hline
Offloaded task size $B^I_k$ & \multicolumn{3}{c|}{(3 Mbits, 6 Mbits)} \\
\hline
Output to input bits proportion $\alpha_k$ & \multicolumn{3}{c|}{0.2 \cite{mec2}} \\
\hline
CPU cycles per bit $\beta_k$ & \multicolumn{3}{c|}{330, 960, or 1900 \cite{mec18}} \\
\hline
$M_k$ & 2 & N/A &N/A \\
\hline
$\nu_m$ & N/A & 2 & 0.8 \\
\hline
\end{tabular}
\end{table}

\begin{figure}[!ht]
\centerline{\includegraphics[width=8cm]{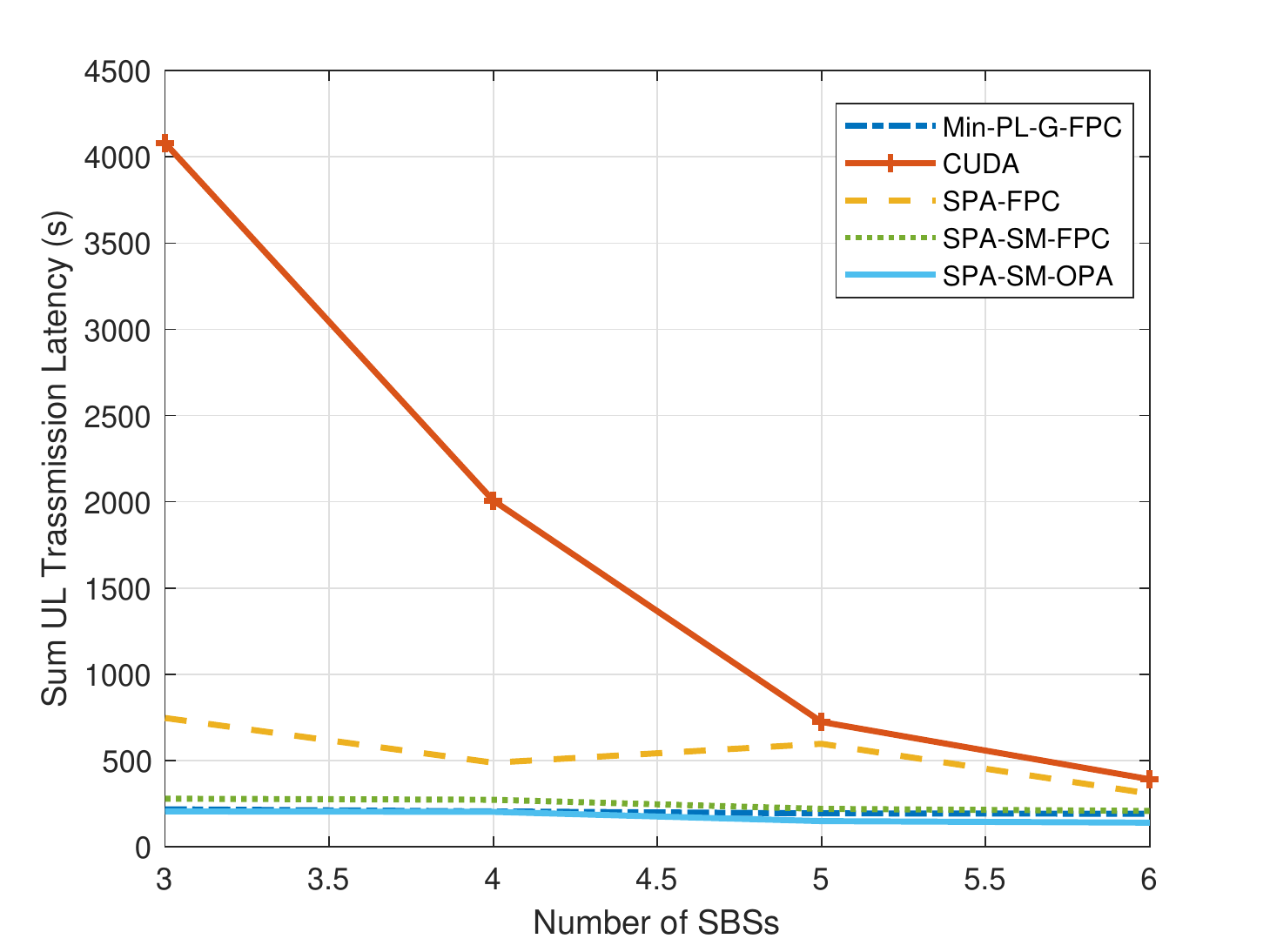}}\vspace{-0.1in}
\caption{UL sum transmission latency.}\label{Fig3}
\end{figure}

\begin{figure}[!ht]
\centerline{\includegraphics[width=8cm]{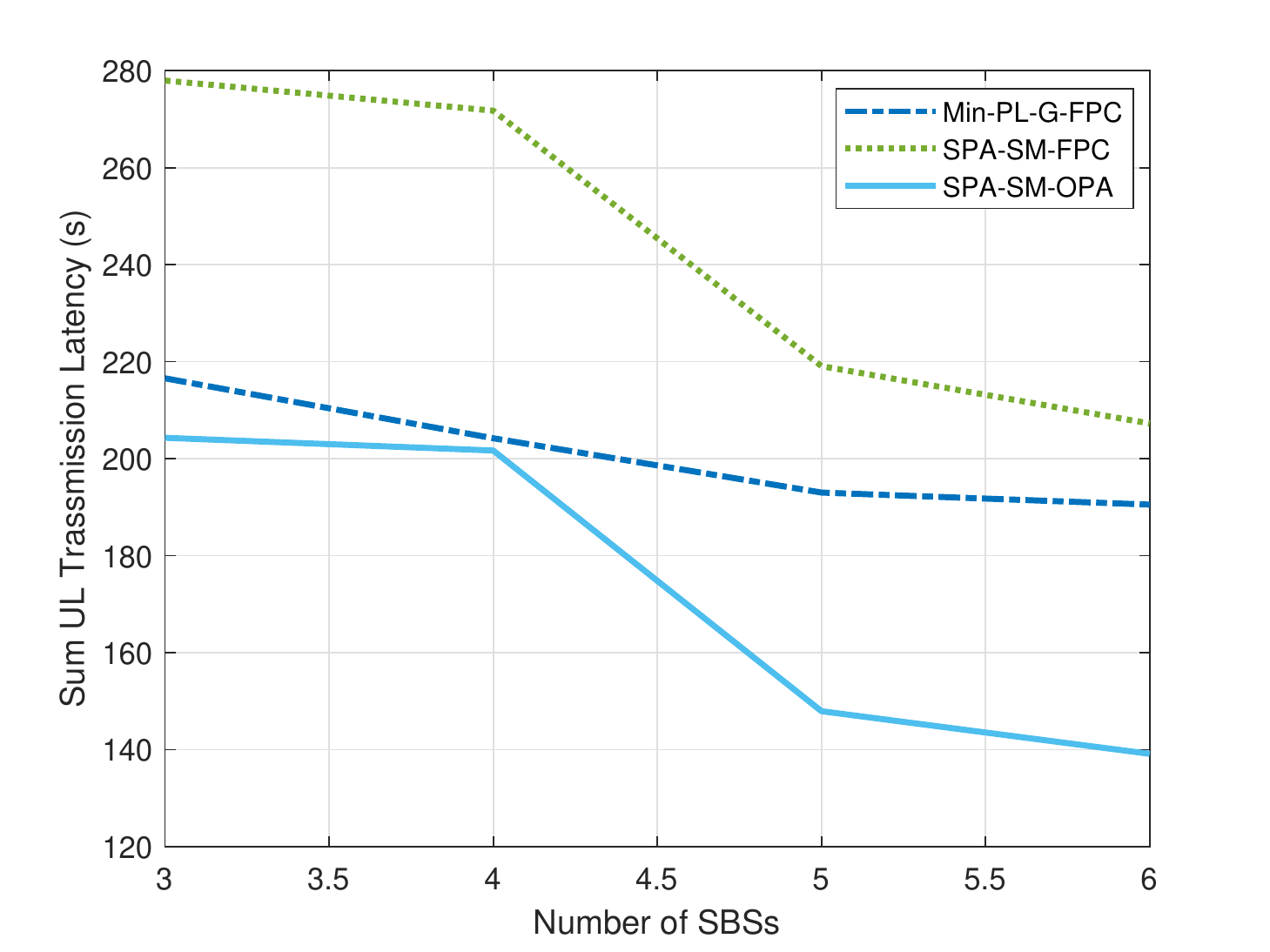}}\vspace{-0.075in}
\caption{UL sum transmission latency.}\label{Fig4}
\end{figure}

Fig. \ref{Fig3} depicts the relationship between the UL sum transmission latency and the number of SBSs. As can be seen, the UL transmission latency of the CUDA scheme is several times higher than all its decoupled counterparts. This is because the decoupled schemes shorten the MD-BS distance, and reduce the interference in the network. Due to peer effects, the sum-latency of the SPA scheme decreases with the number of SBSs, which demonstrates the importance of swap matching. Since the latency of the CUDA and SPA-FPC schemes are significantly higher than other schemes, Fig. \ref{Fig4} focuses on the Min-PL-G-FPC, SPA-SM-FPC, and SPA-SM-OPA schemes. Particularly, the UL transmission latency of the SPA-SM-FPC is higher than the Min-PL scheme, since some computation-intensive tasks are offloaded to the MBS cloudlet to reduce the computation latency, but the communication latency increases at the same time. Furthermore, with the aid of the proposed OPA scheme, the transmission latency of SPA-SM-OPA decreases by as much as 20\% and is lower than the latency of the Min-PL-G-FPC scheme.

\begin{figure}[htbp]
\centerline{\includegraphics[width=8cm]{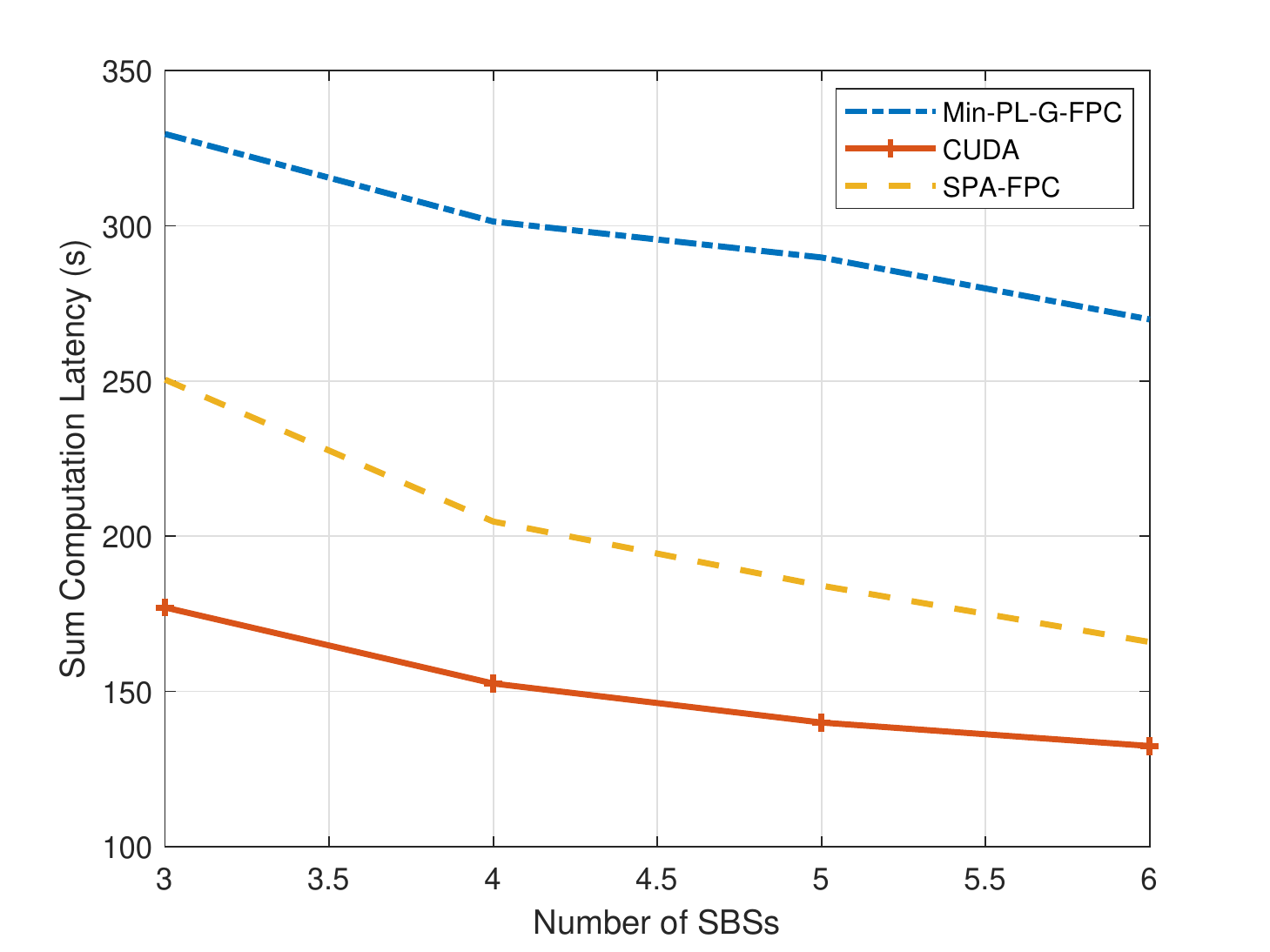}} \vspace{-0.075in}
\caption{Sum computation latency of Min-PL-G-FPC, CUDA and SPA-FPC schemes.}\label{Fig5}
\end{figure}

\begin{figure}[htbp]
\centerline{\includegraphics[width=8cm]{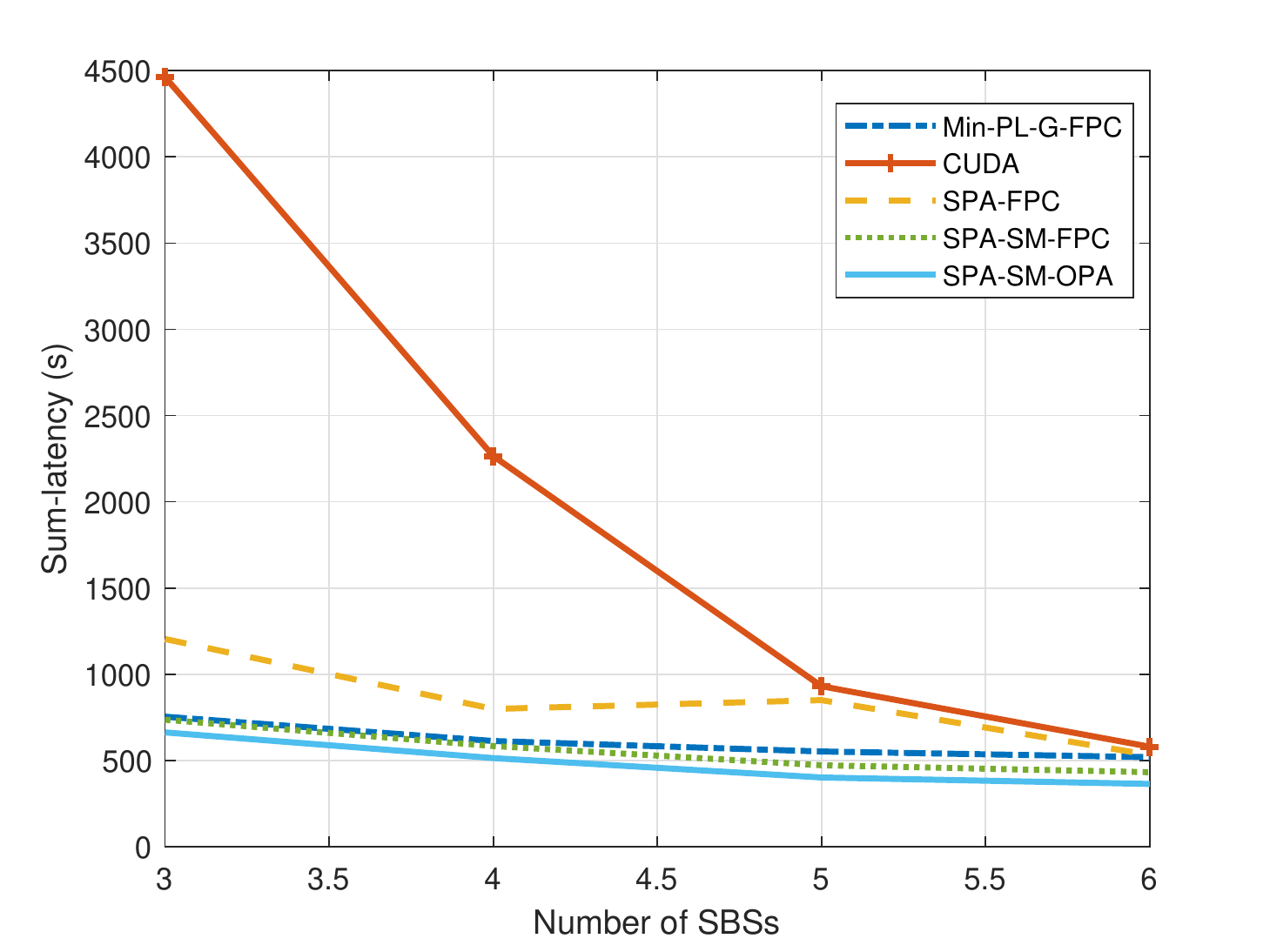}}\vspace{-0.075in}
\caption{Sum-latency.}\label{Fig6}
\end{figure}

Fig. \ref{Fig5} illustrates the sum computation latency of the different schemes. Since the SPA-FPC, SPA-SM-FPC, and SPA-SM-OPA schemes choose the same UL serving BS for each MD, and have the same computation latency, only the SPA-FPC scheme is considered. When the number of SBSs increases, there are more computational resources, and so the computation latency decreases. It can be seen that the SPA-FPC scheme has the lowest computation latency, because most computation-intensive tasks are offloaded to the MBS cloudlet that has high computation capability. Most of the MDs are associated with the MBS by the CUDA scheme, and so the computation latency for the CUDA scheme is also low. As for the Min-PL-G-FPC scheme, it offloads most MDs to the SBSs and does not offload all the computation-intensive tasks to the MBS cloudlet, thus its computation latency is the highest. 

\begin{figure}[htbp]
\centerline{\includegraphics[width=8cm]{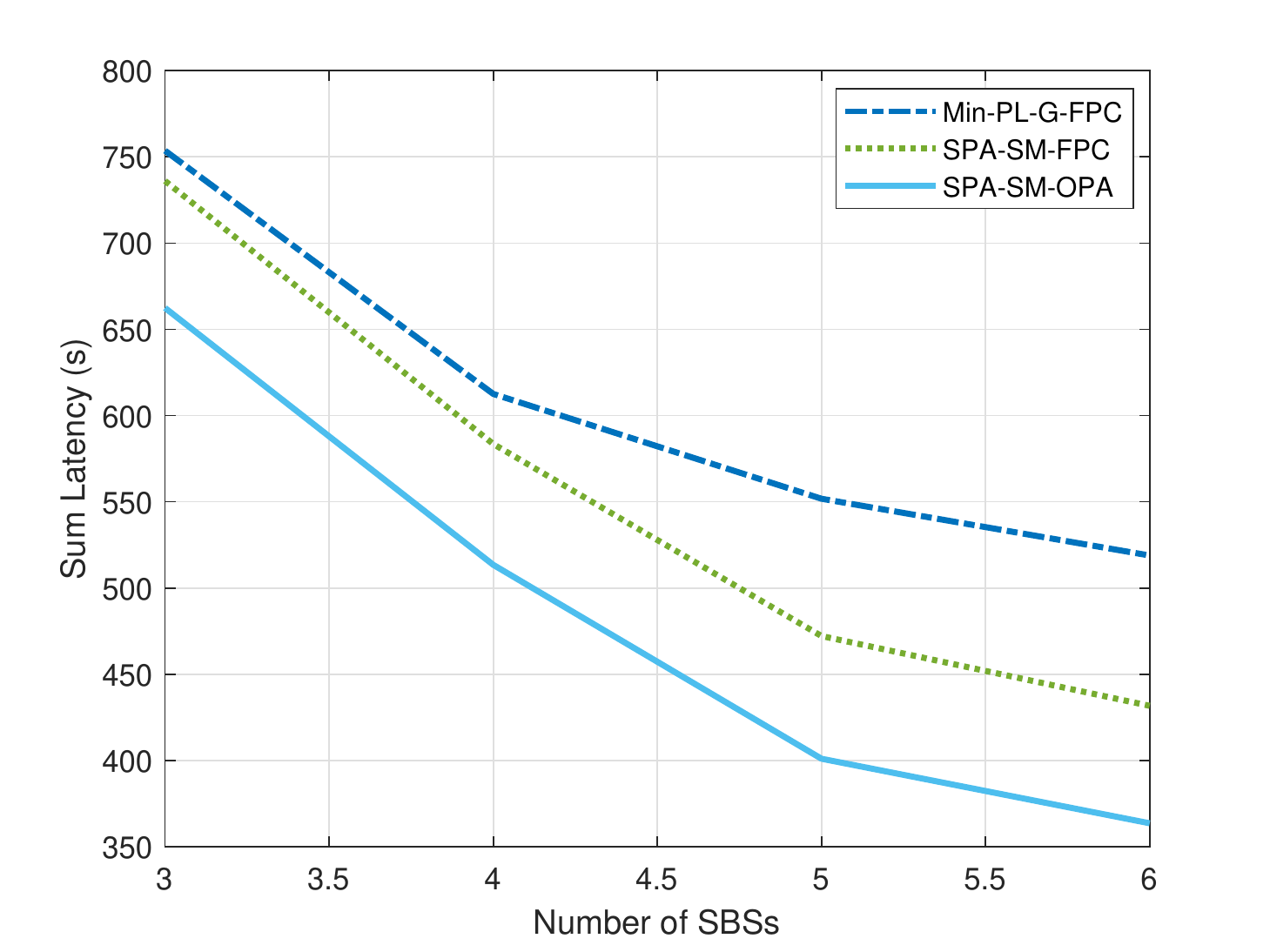}}\vspace{-0.075in}
\caption{Sum-latency of Min-PL-G-FPC, SPA-SM-FPC and SPA-SM-OPA schemes.}\label{Fig7}
\end{figure}

Fig. \ref{Fig6} illustrates the sum-latency of all schemes, which is the sum of UL and DL transmission latency, backhaul latency, and computation latency. As the DL latency of all the schemes is the same and the backhaul latency is very small compared to communication and computation latency, the sum-latency mainly depends on the UL transmission latency and computation latency. It is clear from Fig. \ref{Fig6} that the latency of all the DUDe schemes is much lower than that of the CUDA scheme. The sum-latency of the SPA-SM-OPA scheme is the lowest, and only 15\% to 60\% of the CUDA scheme. According to Fig. \ref{Fig7}, the sum-latency of SPA-SM-OPA scheme is around 15\% lower than that of the Min-PL-G-FPC scheme.

\begin{figure}[htbp]
\centerline{\includegraphics[width=8cm]{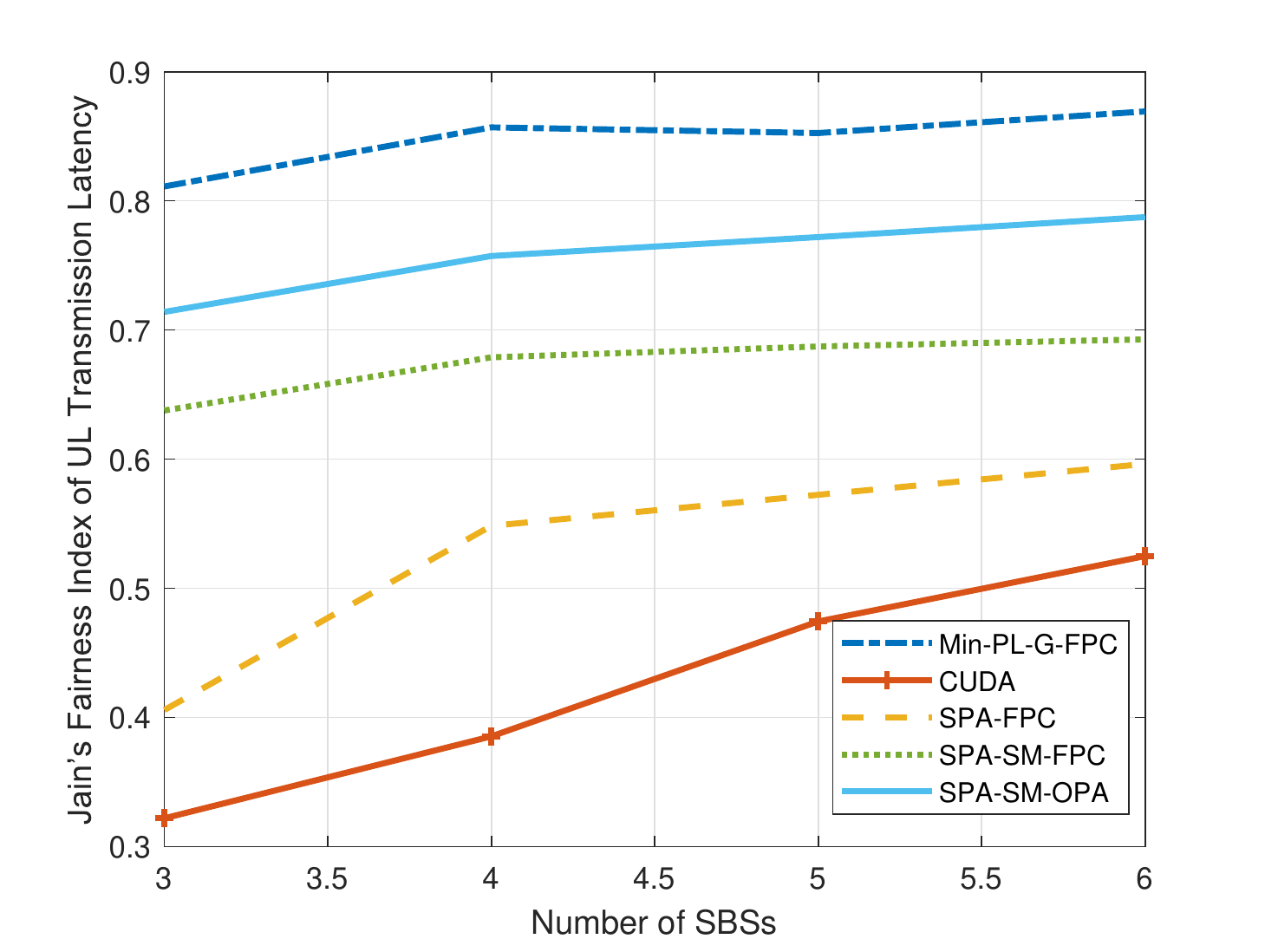}}\vspace{-0.075in}
\caption{Jain’s fairness index of UL transmission latency vs. number of SBSs.}\label{Fig9}
\end{figure}

\begin{figure}[htbp]
\centerline{\includegraphics[width=8cm]{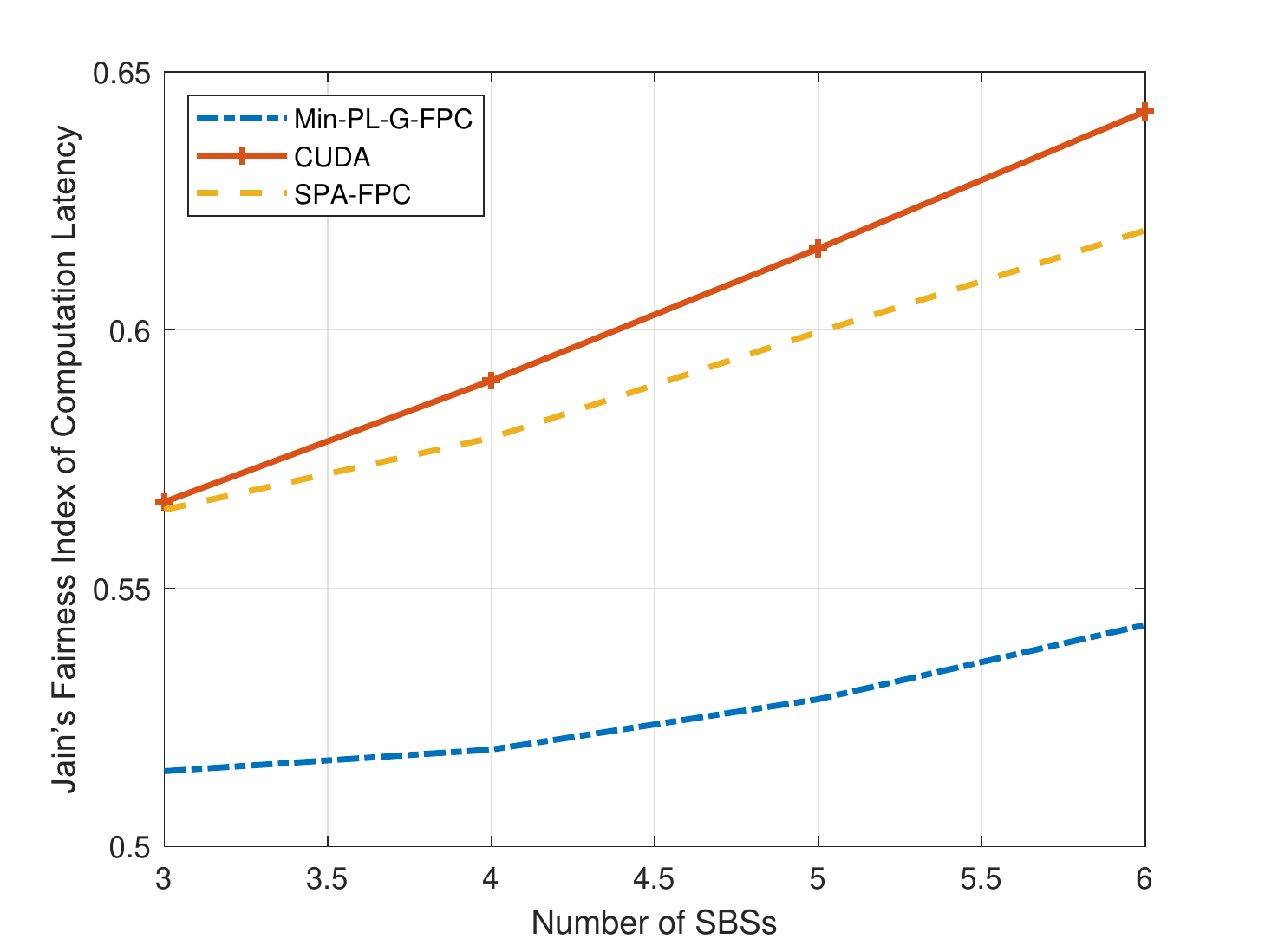}}\vspace{-0.075in}
\caption{Jain’s fairness index of computation latency vs. number of SBSs.}\label{Fig10}
\end{figure}

Another important metric to consider in this scenario is the fairness metric. Figs. \ref{Fig9} and \ref{Fig10} utilize Jain’s fairness index to evaluate the UL transmission latency and computation latency fairness, respectively \cite{jain1984quantitative}. As the BSs are uniformly distributed in our simulations, the Min-PL scheme---which selects the UL serving BS from the perspective of the UL path-loss---yields the highest UL transmission fairness, as can be seen from Fig. \ref{Fig9}. As the SPA scheme jointly considers the computation and transmission latency, and offloads the computationally-intensive tasks to the MBS, it may cause slightly higher transmission latency to those MDs. In turn, the fairness is marginally reduced, but is still higher than the traditional CUDA scheme, especially when combining SPA with swap matching, which helps reduce the inter-cell interference. As for the computation latency fairness shown in \ref{Fig10}, the fairness index of CUDA scheme appears to be the highest as the CUDA scheme associates most the MDs to the MBS, and the rest to the SBS, which have limited computation resource. The SPA scheme allocates the computationally-intensive tasks to the MBSs and the others to the nearby SBSs, which can also achieve similar fairness as the CUDA scheme{\footnote{SPA-FPC, SPA-SM-FPC and SPA-SM-OPA schemes have the same computation latency, as the MD-BS pairs are the same in these schemes.}}. The Min-PL scheme does not consider the computation capabilities of different kinds of BSs, hence the MDs which have computationally-intensive tasks and happen to be near the SBSs will suffer from high computation latency, thus its fairness index is the lowest.                           

\begin{figure}[htbp]
    \centerline{\includegraphics[width=8cm]{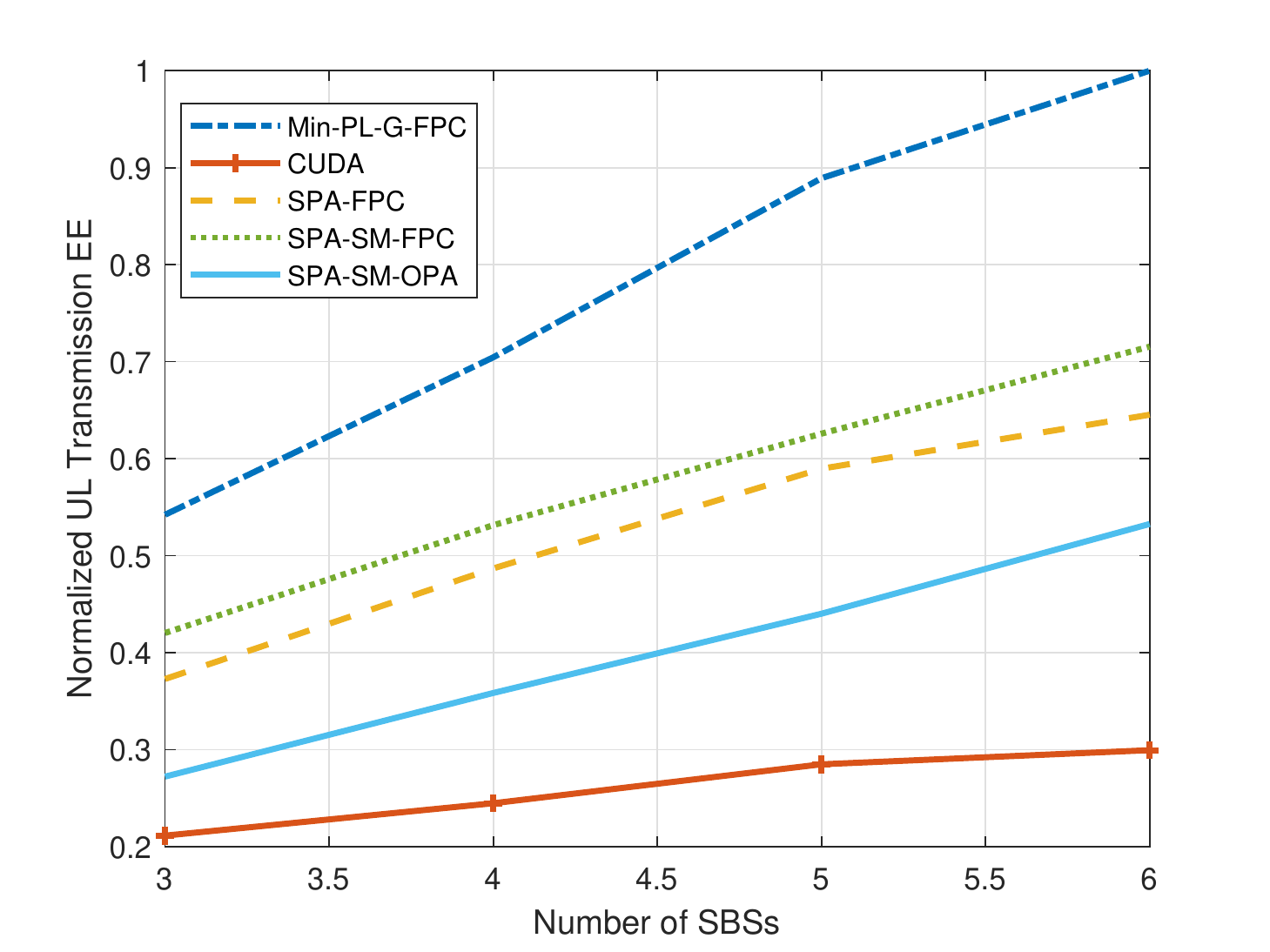}}\vspace{-0.075in}
    \caption{Normalized UL transmission EE vs. number of SBSs.}\label{Fig8}
\end{figure}

Fig. \ref{Fig8} depicts the UL EE of different schemes. It can be seen that the normalized EE of the Min-PL-G-FPC scheme is the highest, since it associates each MD with their nearest BS; while the EE of the CUDA scheme is the lowest, which associates most MDs to the MBS. Furthermore, swap matching can improve the SINR of some MDs by swapping their subchannels, and thus improves their EE. The EE of the SPA-SM-FPC scheme is twice higher than the Min-PL-G-FPC scheme. The aim of the proposed power allocation scheme is to reduce the network sum-latency, and so the transmit powers of some MDs are increased, which reduces EE.

\begin{figure}[htbp]
\centerline{\includegraphics[width=9cm]{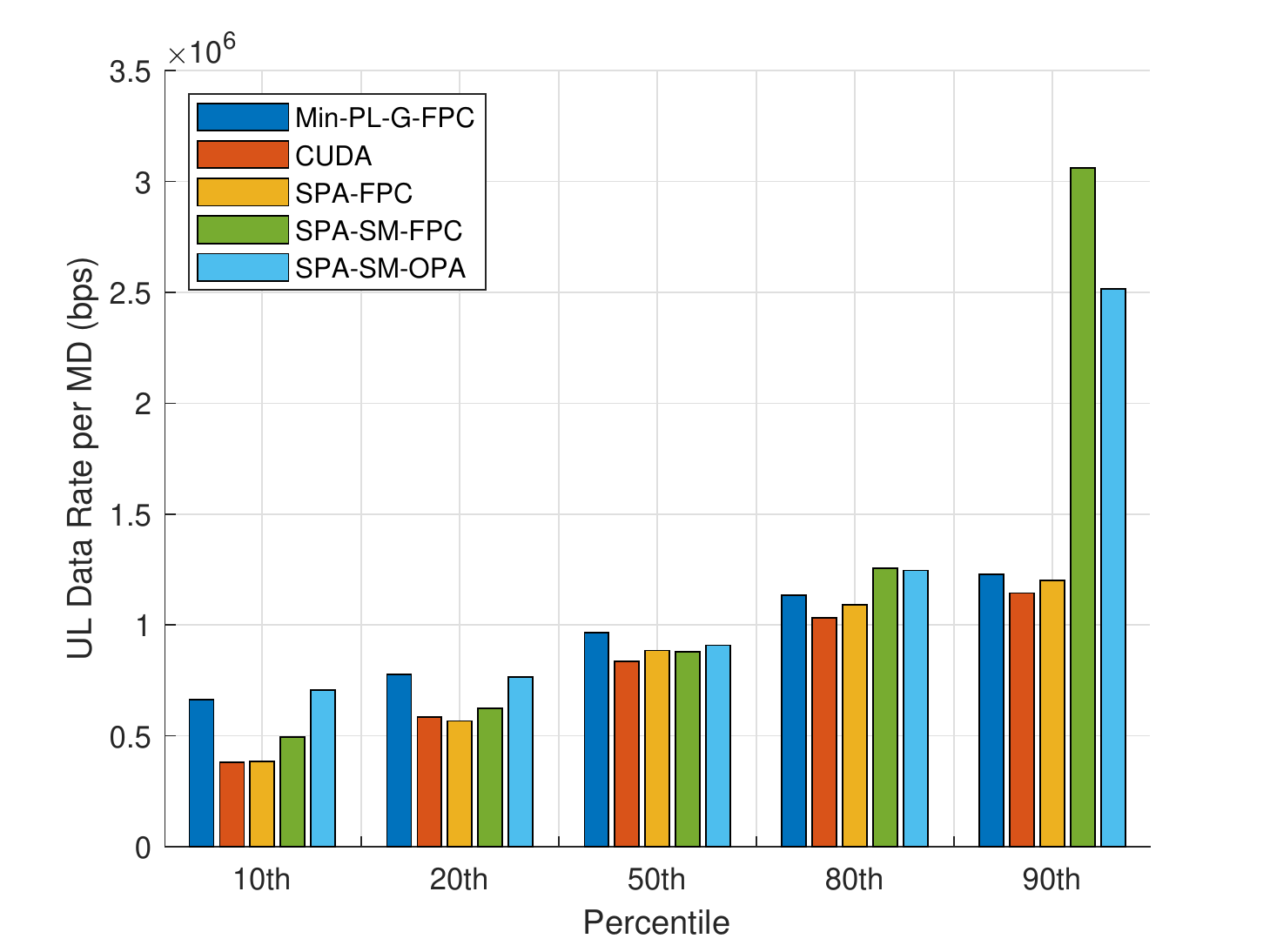}}\vspace{-0.075in}
\caption{$10^{th}$, $20^{th}$, $50^{th}$, $80^{th}$, and $90^{th}$ percentile MD data rates based on different schemes.}\label{Fig12}
\end{figure}

Fig. \ref{Fig12} depicts the $10^{th}$, $20^{th}$, $50^{th}$, $80^{th}$, and $90^{th}$ percentile data rates per MD. It can be seen that the CUDA scheme has the lowest data rates, especially the $10^{th}$ percentile data rates, because it allocates too many MDs to the MBS; and for those MDs at the MBS edge, their UL data rates are relatively low. The Min-PL scheme selects the UL serving BS by the UL path-loss, which results in a balanced MD distribution among BSs, and the data rate distribution is the most balanced (e.g. the $90^{th}$ percentile data rate of the Min-PL scheme is only 70\% higher than it $10^{th}$ percentile data rate). The SPA-SM scheme has lower $10^{th}$, $20^{th}$, $50^{th}$ percentile data rates as it offloads more MDs to the MBS. However, its $80^{th}$ and $90^{th}$ percentile data rates are higher, which is due to its better inter-cell interference control. Compared with the SPA-SPA-SM-FPC scheme, the proposed optimal power allocation scheme (OPA) increases the $10^{th}$, $20^{th}$, $50^{th}$ percentile data rates at the expense of the $80^{th}$ and $90^{th}$ percentile data rates, which reduces the overall UL transmission latency, as seen in Fig. \ref{Fig4}.

\section{Conclusion}\label{conclusion}
This paper has explored the utilization of DUDe in MEC networks. Existing research works considered BS association and subchannel allocation separately, and studied MEC under UL/DL coupled single BS association. In contrast, this paper has provided a new perspective on resource allocation in MEC-enabled heterogeneous networks. Our results have successfully demonstrated that the latency of all the DUDe schemes are much lower than that of the CUDA scheme. Specifically, the network sum-latency of the SPA-SM-OPA scheme is the lowest among all the DUDe schemes, which is only 15\% to 60\% of the CUDA scheme, and is around 15\% lower than that of the Min-PL-G-FPC scheme. The fairness of the UL transmission latency of the SPA-SM-OPA scheme is also higher than that of the CUDA scheme. Furthermore, the EE of the SPA-SM-FPC scheme is two times higher than the CUDA scheme. The users' data rates under the proposed scheme are increased as well, especially for those cell edge MDs. Notwithstanding, some limitations must not be overlooked. For example, the proposed schemes are centralized in nature hence require a control node, hence if there is no such node in the network, then distributed alternatives must be sought, which could be the subject of further research in the future.

\vspace{-0.05in}
\section*{Acknowledgement}\label{acknowledgement}
This work was partially supported by the Kuwait Foundation for the Advancement of Sciences (KFAS) under project code PN17-15EE-02.

\bibliographystyle{IEEEtran}
\bibliography{library}
\end{document}